\documentclass[preprint,fleqn,11pt]{elsarticle}

\usepackage{amsthm} 
\usepackage{amssymb} 
\usepackage{amstext} 
\usepackage{amsmath} 
\usepackage{bm} 
\usepackage{graphicx} 
\usepackage{epstopdf} 
\usepackage{subcaption} 
\usepackage{float}
\usepackage{tikz}
\usepackage{lineno,hyperref}
\usepackage[capitalize]{cleveref} 
\usepackage[makeroom]{cancel} 
\usepackage{multirow} 
\usepackage{booktabs} 
\usepackage{lscape}
\usepackage{tabularx} 
\usepackage{adjustbox}
\usepackage{tablefootnote}
\usepackage{threeparttablex}
\usepackage{longtable}
\usepackage{arydshln}
\makeatletter
\newcommand{\mathleft}{\@fleqntrue\@mathmargin0pt}
\newcommand{\mathcenter}{\@fleqnfalse}
\makeatother

\modulolinenumbers[5]

\crefalias{subequation}{equation} 
\crefalias{eqnarray}{equation} 
\crefformat{pluraleq}{Eqs.~(#2#1#3)} 
\Crefformat{pluraleq}{Equations~(#2#1#3)} 

\newcommand{\mr}{RM$^{\dagger}$-CN }
\newcommand{\lpm}{\resizebox{0.025\hsize}{!}{$+/-$}}
\newcommand{\lmp}{\resizebox{0.025\hsize}{!}{$-/+$}}
\newcommand{\SN}{S$_{\text{N}}$ }
\def\bal#1\nal{\begin{align}#1\end{align}}
\def\bala#1\nala{\begin{align*}#1\end{align*}}
\def\bsub#1\nsub{\begin{subequations}#1\end{subequations}}

\journal{ }

 \biboptions{sort&compress}

\begin{document}

\begin{frontmatter}

\title{On the calculation of neutron sources generating steady prescribed power distributions in subcritical systems using multigroup X,Y-geometry discrete ordinates models 
}\author[iprj]{L.R.C. Moraes\corref{cor1}}
\author[iprj]{H. Alves Filho\fnref{hermes}}
\author[iprj]{R. C. Barros\fnref{barros}}

\address[iprj]{Instituto Politécnico,  IPRJ/UERJ, P.O.Box 97282, 28610-974, Nova Friburgo, RJ, Brasil}
\cortext[cor1]{Corresponding author:  leonardrcmoraes@gmail.com}
\fntext[hermes]{halves@iprj.uerj.br}
\fntext[barros]{ricardo.barros@uerj.br}

\begin{abstract}
In this paper a methodology is described to estimate multigroup neutron source distributions which must be added into a subcritical system to drive it to a steady state prescribed power distribution. This work has been motivated by the principle of operation of the ADS (Accelerator Driven System) reactors, which have subcritical cores stabilized by the action of external sources. We use the energy multigroup two-dimensional neutron transport equation in the discrete ordinates formulation (S$_{\text{N}}$) and the equation which is adjoint to it, whose solution is interpreted here as a distribution measuring the importance of the angular flux of neutrons to a linear functional. These equations are correlated through a reciprocity relation, leading to a relationship between the interior sources of neutrons and the power produced by unit length of height of the domain. A coarse-mesh numerical method of the spectral nodal class, referred to as adjoint response matrix constant-nodal method, is applied to numerically solve the adjoint \SN equations. Numerical experiments are performed to analyze the accuracy of the present methodology so as to illustrate its potential practical applications.
\end{abstract}

\end{frontmatter}

\section{Introduction}
	Although the concept of the ADS reactor (Accelerator Driven System) appears to be new, since it is classified as a Generation IV reactor \cite{Gen4}, in fact similar ideas were first proposed more than 50  years ago \cite{Nifenecker,Nifenecker2001}. At that time they were not pushed forward though, mainly due to the lack of economic investment as a result of their complexities in comparison with other reactors proposed at that time, such as the light-water reactors: PWR (Pressurized Water Reactor) and BWR (Boiling Water Reactor). However, radioactive waste concerns involving light-water reactors in conjunction to greenhouse effects have contributed to put the ADS reactor back to the game \cite{Akkaya2018,Kapoor2002,Sakon2013,Gulik2014,Chen2015,AitAbderrahim2012}.

The ADS reactor, differently from commercial nuclear reactors of the present date, possesses a subcritical core. This subcritical core is stabilized by a stationary source of neutrons which is normally generated by the collision of high-energy protons with the atomic nuclei of heavy metals placed within the reactor core. This phenomenon is termed nuclear spallation and the protons are accelerated by an accelerator which operates with a fraction of the power generated by the reactor itself \cite{Lisowski1990,Spallation,Bauer2001}. As the subcritical core is sustained by the particle accelerator only, one just needs to switch off the accelerator to shut down the reactor. Therefore, the ADS reactor is viewed as an intrinsically safe device. Moreover, it might offer interesting transmutation possibilities and be a smart solution to incinerate heavy actinides.

Motivated by the principle of operation of the ADS reactor, we describe in this work a methodology to determine the neutron source distribution required to drive a subcritical system to a prescribed steady-state distribution of power. To represent the physical phenomena used as bases for the present methodology, we consider the energy multigroup X,Y-geometry discrete ordinates (S$_{\text{N}}$) mathematical model. It is well known that the solution of the equation which is adjoint to the neutron transport equation can be interpreted as an importance function \cite{Ga67,Gandini87,Lewins65,HB10}, which allows its application in a variety of applications, such as the described in this paper. The methodology is centered on the definition of a specific matrix, called importance matrix, which correlates explicitly the neutron source distribution and the power densities generated by the fuel regions of the subcritical system, as in two-dimensional calculations, nuclear power is determined per unit length of height of the domain. The importance matrix is composed of solutions of the energy multigroup adjoint \SN equations that represent, in the context of this work, a measure of the importance that one neutron, inserted into the system, has to power generation. 

 In \cref{sec2} we describe the methodology we use to determine the required neutron source distribution.  As the importance matrix is composed by solutions of the multigroup adjoint S$_{\text{N}}$ equations, we present in \cref{sec3} the coarse-mesh adjoint response matrix constant nodal (RM$^{\dagger}$-CN) method \cite{Moraes20} which is used in conjunction to the adjoint partial one-node block inversion iterative scheme to generate numerical solutions for the adjoint angular fluxes on coarse spatial grids. In \cref{sec4} we present numerical experiments for two model problems in order to analyze the accuracy of the present methodology and in \cref{sec5} we provide a number of general concluding remarks and suggestions for future work.

\section{Methodology}\label{sec2}
\begin{figure}[]
	\centering
	\includegraphics[scale=0.28]{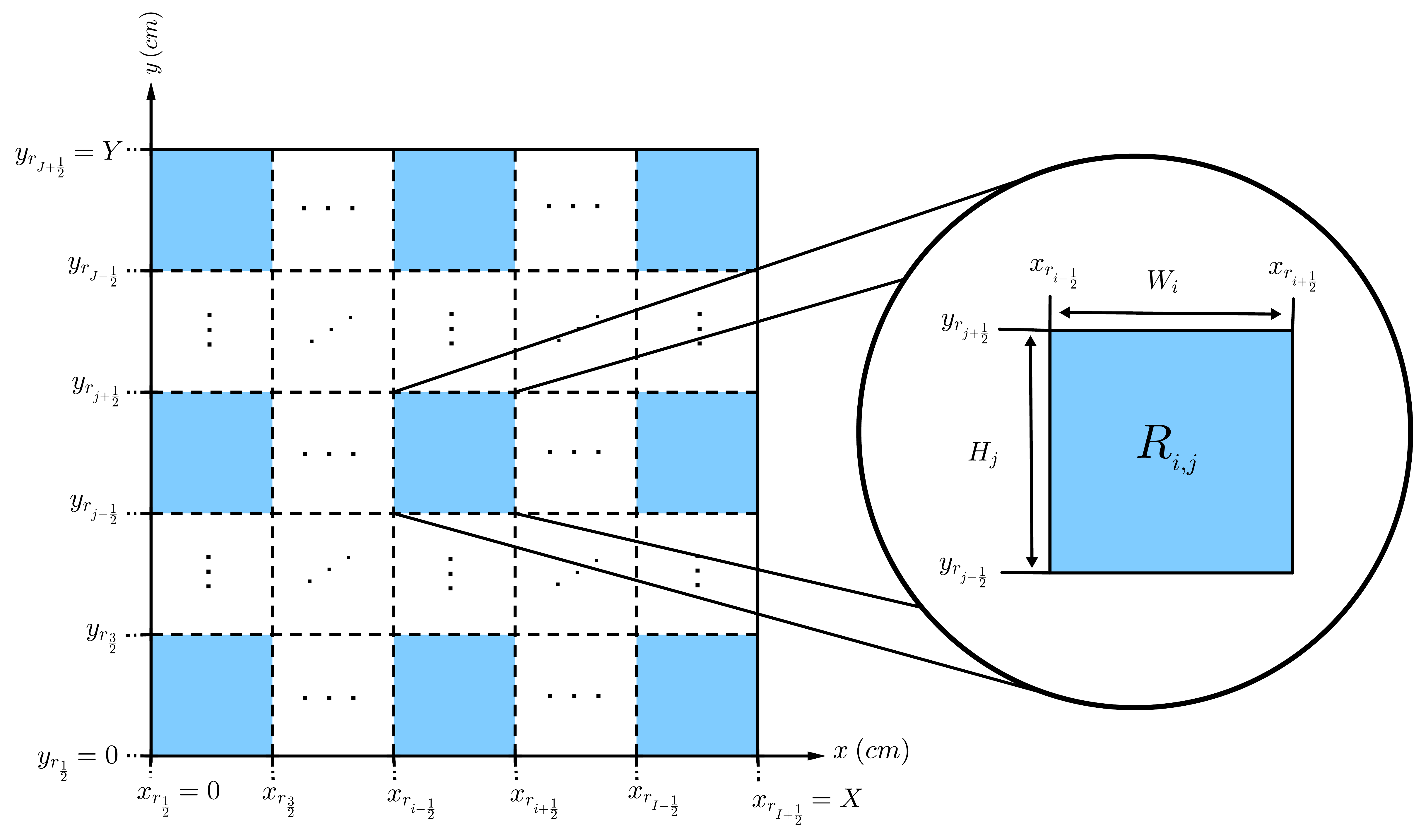}\vspace*{0cm}
	\caption{Two-dimensional spatial domain with $IJ$ contiguous regions $R_{i,j}$ of width $W_{_{i}}$ and height $H_{_{j}}$.}
	\label{f1}
\end{figure}

Let us consider the functional
\begin{equation}\label{e2.1}
F = \int_{0}^{Y}\int_{0}^{X}\sum_{g=1}^{G}\sum_{m=1}^{M}Q^{\dagger}_{_{g}}(x,y)\Psi_{_{g}}(x,y,\mu_{_{m}},\eta_{_{m}})\tau{_{m}}\,dxdy \equiv \left\langle Q^{\dagger},\Psi \right\rangle,
\end{equation} 
which is linear with respect to both the adjoint source term $Q^{\dagger}$ and the neutron angular flux $\Psi$. The group neutron angular flux in \cref{e2.1} satisfies the energy multigroup S$_{\text{N}}$ transport equation \cite{LeMi93} within a rectangle of width $X$ and height $Y$
\begin{subequations}
	\begin{equation}\label{e2.2a}
	T_{_{m,g}}\Psi_{_{g}}(x,y,\mu_{_{m}},\eta_{_{m}}) = Q_{_{g}}(x,y),
	\end{equation}
	where $Q$ the neutron source and
	\begin{equation}
	T_{_{m,g}} = \mu_{_{m}}\frac{\partial}{\partial x}(\cdot)+\eta_{_{m}}\frac{\partial}{\partial y}(\cdot) + \sigma_{t_{i,j,g}}(\cdot) - \frac{1}{4}\sum_{g'=1}^{G}\left(\sigma_{s_{i,j,g'\rightarrow g}} + \chi_{_{i,j,g}}\nu\sigma_{f_{i,j,g'}} \right)\sum_{n=1}^{M}(\cdot)\tau_{_{n}}, \label{e2.2b}
	\end{equation}
	for $m=1:M$, $g=1:G$, $(x,y)\in R_{_{i,j}}$. 
	
	Equation (\ref{e2.2a}) is subject to the boundary conditions, in accordance with \cref{f1},
	\begin{equation}\label{e2.2c}
	\Psi_{_{g}}(0,y,\mu_{_{m}},\eta_{_{m}}) = \lambda_{_{l}}\Psi_{_{g}}(0,y,-\mu_{_{m}},\eta_{_{m}}),\:\mu_{_{m}}>0,
	\end{equation}
	\begin{equation}\label{e2.2d}
\Psi_{_{g}}(X,y,\mu_{_{m}},\eta_{_{m}}) = \lambda_{_{r}}\Psi_{_{g}}(X,y,-\mu_{_{m}},\eta_{_{m}}),\:\mu_{_{m}}<0,
	\end{equation}
	\begin{equation}\label{e2.2e}
\Psi_{_{g}}(x,0,\mu_{_{m}},\eta_{_{m}}) = \lambda_{_{b}}\Psi_{_{g}}(x,0,\mu_{_{m}},-\eta_{_{m}}),\:\eta_{_{m}}>0
	\end{equation}
and
	\begin{equation}\label{e2.2f}
\Psi_{_{g}}(x,Y,\mu_{_{m}},\eta_{_{m}}) = \lambda_{_{t}}\Psi_{_{g}}(x,Y,\mu_{_{m}},-\eta_{_{m}}),\:\eta_{_{m}}<0,
\end{equation}
where $\lambda_{_{u}} = 0$ (vacuum) or $\lambda_{_{u}}=1$ (reflective) indicates the type of boundary condition that each contour of the two-dimensional rectangular domain is subjected to ($u = l\text{ and r; t\text{ and }b}$). In \cref{e2.1,2.2}, $\mu_{_{m}},\,\eta_{_{m}}$ are discrete directions of motion and $\tau_{_{m}}$ are the corresponding weights of the angular quadrature; in our case, the Level Symmetric Quadrature (LQ$_{\text{N}}$) \cite{LeMi93}, where N is the order of the discrete ordinates formulation (S$_{\text{N}}$) and $M= \frac{\text{N}\left( \text{N}+2\right) }{2}$ is the total number of discrete directions. $G$ is the number of energy groups in the multigroup formulation \cite{BeGl70}. Moreover, the quantities  $\sigma_{t}$, $\sigma_{s}$ and $\sigma_{f}$ are the total, scattering and fission macroscopic cross sections, with $\nu$ and $\chi$ being respectively the average number of neutrons released in a fission event and fission spectrum. The two-dimensional domain considered in \cref{e2.1,2.2} are composed by $IJ$ contiguous regions $R_{_{i,j}}$, with $i=1:I$ and $j=1:J$, viz \cref{f1}, where we have considered material parameters and source terms as uniform with respect to the spatial variables inside each region $R_{_{i,j}}$.
	\label[pluraleq]{2.2}
\end{subequations}

Now, in order to analyze the definition of the functional, as presented in \cref{e2.1}, let us suppose we introduce one neutron inside the rectangular domain, migrating in a specific direction in a given energy group. Due to this inserted neutron, the angular flux will experiment an increase $\Delta \Psi$. This increase in the neutron angular flux consequently promotes a change $\Delta F$ in the functional defined as neutron importance. The neutron importance, also termed adjoint angular flux in this work, quantifies the contribution (change) $\Delta F$ to the functional by the addition of one neutron inside the system. This contribution may be produced either directly by this inserted neutron or, in presence of a multiplying system, through its precursor \cite{Lewins65,Gandini87}.

Making use of the neutron importance concept we can rewrite the functional defined in \cref{e2.1} counting the contribution of each neutron introduced in the system by the neutron source. We note that neutrons entering the system through the boundaries (reflective boundary condition) are composed directly or indirectly (through their precursors) of neutrons introduced by the neutron source. Thus, their contributions are already included by the source term. Therefore, the functional $F$ can alternatively be expressed as
\begin{equation}\label{e2.3}
F = \int_{0}^{Y}\int_{0}^{X}\sum_{g=1}^{G}\sum_{m=1}^{M}Q_{_{g}}(x,y)\Psi^{\dagger}_{_{g}}(x,y,\mu_{_{m}},\eta_{_{m}})\tau_{_{m}}dxdy \equiv \left\langle Q,\Psi^{\dagger} \right\rangle,
\end{equation}
where $\Psi^{\dagger}$ represents the neutron importance, that is, the contribution to the functional of a neutron introduced in the system and $Q$ biases this contribution by all neutrons introduced by the source. Considering \cref{e2.1,e2.3} we obtain
\begin{equation}\label{e2.4}
\left\langle Q^{\dagger},\Psi \right\rangle  = \left\langle Q,\Psi^{\dagger} \right\rangle,
\end{equation}
which is the well-known source reciprocity relation \cite{Gandini87}. Observing \cref{e2.1,e2.2a,e2.3} we may assume that $Q^{\dagger}$ is the source term of the equation whose solution is $\Psi^{\dagger}$. Thus,
\begin{subequations}
	\begin{equation}\label{e2.5a}
	T_{_{m,g}}^{\dagger}\Psi^{\dagger}_{_{g}}(x,y,\mu_{_{m}},\eta_{_{m}}) = Q^{\dagger}_{_{g}}(x,y),
	\end{equation}
	where $m=1:M$, $g=1:G$ and $(x,y)\in R_{_{i,j}}$. Equation (\ref{e2.5a}) is referred to as the energy multigroup adjoint S$_{\text{N}}$ equations. At this point we remark that the importance function $\Psi^{\dagger}$ is also referred to as adjoint angular flux, since it is the solution of the energy multigroup adjoint S$_{\text{N}}$ equations.

	 Substituting \cref{e2.2a,e2.5a} into \cref{e2.4} and following an analogous procedure, as described in reference \cite{HB10}, we obtain the operator
	\begin{equation}\label{e2.5b}
	T_{_{m,g}}^{\dagger} = -\mu_{_{m}}\frac{\partial}{\partial x}(\cdot) -\eta_{_{m}}\frac{\partial}{\partial y}(\cdot) + \sigma_{t_{i,j,g}}(\cdot) - \frac{1}{4}\sum_{g'=1}^{G}\left(\sigma_{s_{i,j,g\rightarrow g'}} + \chi_{_{i,j,g'}}\nu\sigma_{f_{i,j,g}} \right)\sum_{n=1}^{M}(\cdot)\tau_{_{n}}.
	\end{equation}
	Equation (\ref{e2.5a}) is subject to zero adjoint angular fluxes in the outgoing directions ($\lambda_{_{u}} = 0$)  or reflective ($\lambda_{_{u}} = 1$) boundary conditions. That is to say,
	\begin{equation}\label{e2.5c}
\Psi^{\dagger}_{g}(0,y,-\mu_{_{m}},\eta_{_{m}}) = \lambda_{_{l}}\Psi^{\dagger}_{_{g}}(0,y,\mu_{_{m}},\eta_{_{m}}),\:\mu_{_{m}}>0,
	\end{equation}
		\begin{equation}\label{e2.5d}
	\Psi^{\dagger}_{_{g}}(X,y,-\mu_{_{m}},\eta_{_{m}}) = \lambda_{_{r}}\Psi^{\dagger}_{_{g}}(X,y,\mu_{_{m}},\eta_{_{m}}),\:\mu_{_{m}}<0,
	\end{equation}
		\begin{equation}\label{e2.5e}
	\Psi^{\dagger}_{_{g}}(x,0,\mu_{_{m}},-\eta_{_{m}}) = \lambda_{_{b}}\Psi^{\dagger}_{_{g}}(x,0,\mu_{_{m}},\eta_{_{m}}),\:\eta_{_{m}}>0 
	\end{equation}
	and
	\begin{equation}\label{e2.5f}
	\Psi^{\dagger}_{_{g}}(x,Y,\mu_{_{m}},-\eta_{_{m}}) = \lambda_{_{t}}\Psi^{\dagger}_{_{g}}(x,Y,\mu_{_{m}},\eta_{_{m}}),\:\eta_{_{m}}<0.
	\end{equation}
	\label[pluraleq]{2.5}
\end{subequations}

Furthermore, let us focus our attention on the functional presented in \cref{e2.1}. Analyzing this equation we observe that the physical meaning of $F$ and the definition of $Q^{\dagger}$ are directly correlated. In other words, when we assume a specific meaning for the functional, we define consequently the adjoint source term. As we are interested in building a relation between the neutron source and the power generated by the fuel regions, we assume that the functional $F$ represents the power density generated by a fuel region ($P_{_{i^{\star},j^{\star}}}$). Therefore, we write
\begin{equation}\label{e2.6}
F = P_{_{i^{\star},j^{\star}}} = \int_{0}^{Y}\int_{0}^{X}\sum_{g=1}^{G}\sum_{m=1}^{M}\epsilon\,\sigma_{f_{g}}(x,y)\delta_{_{i^{\star},j^{\star}}}(x,y)\Psi_{_{g}}(x,y,\mu_{_{m}},\eta_{_{m}})\tau_{_{m}}dxdy,
\end{equation}
where $\delta_{_{i^{\star},j^{\star}}}(x,y) = \left\lbrace\begin{array}{l}
1,\:\text{if }(x,y)\in R_{_{i^{\star},j^{\star}}} \\0,\:\text{otherwise}
\end{array} \right.$, with $R_{_{i^{\star},j^{\star}}}$ being a fuel region of the system and $\epsilon$ the average energy release in one fission event. From \cref{e2.6} one concludes that the adjoint source term is defined as
\begin{equation}\label{e2.7}
Q^{\dagger}_{_{g}}(x,y) = \epsilon\,\sigma_{f_{g}}(x,y)\delta_{_{i^{\star},j^{\star}}}(x,y).
\end{equation}
Considering \cref{e2.1,e2.3,e2.4,e2.7}, for uniform group neutron sources inside each region $R_{_{i,j}}$, we write
\begin{subequations}
	\begin{equation}\label{e2.8a}
	P_{_{i^{\star},j^{\star}}} = \sum_{j=1}^{J}\sum_{i=1}^{I}\sum_{g=1}^{G}Q_{_{g,i,j}}\overline{\Phi}^{\dagger^{i^{\star},j^{\star}}}_{_{g,i,j}}W_{_{i}}H_{_{j}},
	\end{equation}
	where we have defined the adjoint scalar flux averaged within region $R_{_{i,j}}$, in energy group $g$, as generated by \cref{2.5} with adjoint source term given by \cref{e2.7}, as
	\begin{equation}\label{e2.8b}
	\overline{\Phi}^{\dagger^{i^{\star},j^{\star}}}_{_{g,i,j}} = \frac{1}{W_{_{i}}H_{_{j}}}\int_{y_{r_{j-\frac{1}{2}}}}^{y_{r_{j+\frac{1}{2}}}}\int_{x_{r_{i-\frac{1}{2}}}}^{x_{r_{i+\frac{1}{2}}}}\sum_{m=1}^{M}\Psi^{\dagger}_{_{g}}(x,y,\mu_{_{m}},\eta_{_{m}})\tau_{_{m}}dxdy.
	\end{equation}By varying $i^{\star} = 1:I^{\star}$ and $j^{\star} = 1:J^{\star}$ in \cref{e2.8a}, with $I^{\star}J^{\star}$ being the total number of fuel regions inside the system, we obtain the following matrix equation
	\label[pluraleq]{2.8}
\end{subequations}
\begin{subequations}
	\begin{equation}\label{e2.9a}
	\mathbf{P} = \mathbf{L}^{\dagger}\mathbf{Q},
	\end{equation}
	where $\mathbf{P}$ is a $I^{\star}J^{\star}-$dimensional column matrix composed of the prescribed power density distribution; $\mathbf{Q}$ is a $GIJ-$dimensional column matrix composed by the uniform multigroup neutron sources; and $\mathbf{L}^{\dagger}$ is a $I^{\star}J^{\star} \times GIJ$ matrix that we refer to as importance matrix and is defined as
	\begin{equation}\label{e2.9b}
	\mathbf{L}^{\dagger} = \left[ \begin{array}{cccc}
	\overline{\Phi}^{\dagger^{1,1}}_{_{1,1,1}}W_{_{1}}H_{_{1}} & 	\overline{\Phi}^{\dagger^{1,1}}_{_{2,1,1}}W_{_{1}}H_{_{1}}&\dots&	\overline{\Phi}^{\dagger^{1,1}}_{_{G,I,J}}W_{_{I}}H_{_{J}} \\
	\overline{\Phi}^{\dagger^{2,1}}_{_{1,1,1}}W_{_{1}}H_{_{1}} & 	\overline{\Phi}^{\dagger^{2,1}}_{_{2,1,1}}W_{_{1}}H_{_{1}}&\dots&	\overline{\Phi}^{\dagger^{2,1}}_{_{G,I,J}}W_{_{I}}H_{_{J}} \\
	\vdots & \vdots& \ddots& \vdots
	\\	\overline{\Phi}^{\dagger^{I^{\star},J^{\star}}}_{_{1,1,1}}W_{_{1}}H_{_{1}} & 	\overline{\Phi}^{\dagger^{I^{\star},J^{\star}}}_{_{2,1,1}}W_{_{1}}H_{_{1}}&\dots&	\overline{\Phi}^{\dagger^{I^{\star},J^{\star}}}_{_{G,I,J}}W_{_{I}}H_{_{J}}
	\end{array}\right].
	\end{equation}
	\label[pluraleq]{2.9}
\end{subequations}
\hspace*{-0.15cm}We see that varying $i^{\star} = 1:I^{\star}$ and $j^{\star} = 1:J^{\star}$ in \cref{e2.8a} means solving \cref{2.5} $I^{\star}J^{\star}$ times for each adjoint source as defined in \cref{e2.7}. 

Now we analyze the linear system represented in \cref{e2.9a}. Following linear algebra analysis of \cref{e2.9a}, we obtain the following result: \begin{equation}\label{e2.10}
rank{\left[ \mathbf{L}^{\dagger}\right] } = rank{\left[ \mathbf{L}^{\dagger}|\mathbf{P}\right]} = I^{\star}J^{\star},
\end{equation} where $rank{\left[ \mathbf{L}^{\dagger}\right]}$ stands for the rank of the coefficient matrix and $rank{\left[ \mathbf{L}^{\dagger}|\mathbf{P}\right]}$ stands for the rank of the augmented matrix. Equation (\ref{e2.10}) ensures that the linear system is possible. Moreover, the linear system has a unique solution only when $I^{\star}J^{\star} = GIJ$. Therefore, for monoenergetic problems ($G=1$) composed only of fuel regions ($I^{\star}J^{\star} = IJ$), the neutron source distribution ($\mathbf{Q}$) which must be added inside a subcritical system to drive it to a prescribed power density ($\boldsymbol{\mathbf{P}}$) is directly given by
\begin{equation}\label{e2.11}
\mathbf{Q} = \mathbf{L}^{\dagger^{-1}}\boldsymbol{\mathbf{P}},
\end{equation} 
should $\mathbf{L}^{\dagger}$ have an inverse. However, for most practical problems $G>1$ and $I^{\star}J^{\star}<IJ$. In these cases, the linear system represented in \cref{e2.9a} possesses $GIJ - I^{\star}J^{\star}$ degrees of freedom. Therefore, in order to obtain a unique solution for the problem, one should provide $GIJ - I^{\star}J^{\star}$ auxiliary equations. In fact, the auxiliary equations give to the present methodology a very interesting flexibility, since various choices for auxiliary equations will lead to different possible distributions of neutron sources. Moreover, by fixing the power density distribution, each choice for the set of auxiliary equations should yield a neutron source distribution that, added inside the subcritical system, drives it to the prescribed power density. At this point, we remark that although the auxiliary equations can be chosen freely, not all choices should lead to source distributions that possess coherent physical meaning and we shall illustrate this in \cref{sec4}. 

Another practical situation occurs when the prescribed power density distribution is composed of values of power which are generated by the sum of the power generated by a given set of fuel regions, instead of the power generated by each individual fuel region, as considered in \cref{e2.9a}. To solve this problem we still use the linear system in \cref{e2.9a} rather than solving again the adjoint equations with the appropriate adjoint source terms which would lead to a linear system representing the given power density distribution. As the given value of power is composed of the sum of the power generated by a specific set of fuel regions, we need to sum the rows of  \cref{e2.9a} corresponding to the fuel regions which compose the given value of power. Thus, considering that the prescribed power density is composed of $V$ values of power, we build a $V\times GIJ$ linear system from \cref{e2.9a}, where it is required $GIJ-V$ auxiliary equations in order to obtain a unique solution. Therefore, regardless of the particular numerical experiment, we build \cref{e2.9a} only once and then use it to solve any numerical experiment; hence avoiding to solve S$_{\text{N}}$ adjoint equations for each particular numerical experiment. This is a positive feature of the present technique.

\section{The adjoint response matrix constant-nodal method}\label{sec3}
\setcounter{equation}{0}
Following the methodology described in the previous section, we build the importance matrix in order to determine a neutron source distribution that drives the subcritical system to a prescribed power density distribution. We remark that building the importance matrix is generally time consuming since it requires to solve the multigroup adjoint S$_{\text{N}}$ equations $I^{\star}J^{\star}$ times. Therefore, it is convenient to use the coarse-mesh adjoint response matrix constant-nodal (RM$^{\dagger}$-CN)  method, due to reasons that will become apparent at the end of this section. 

\begin{figure}
 	\centering
 	\includegraphics[scale=0.26]{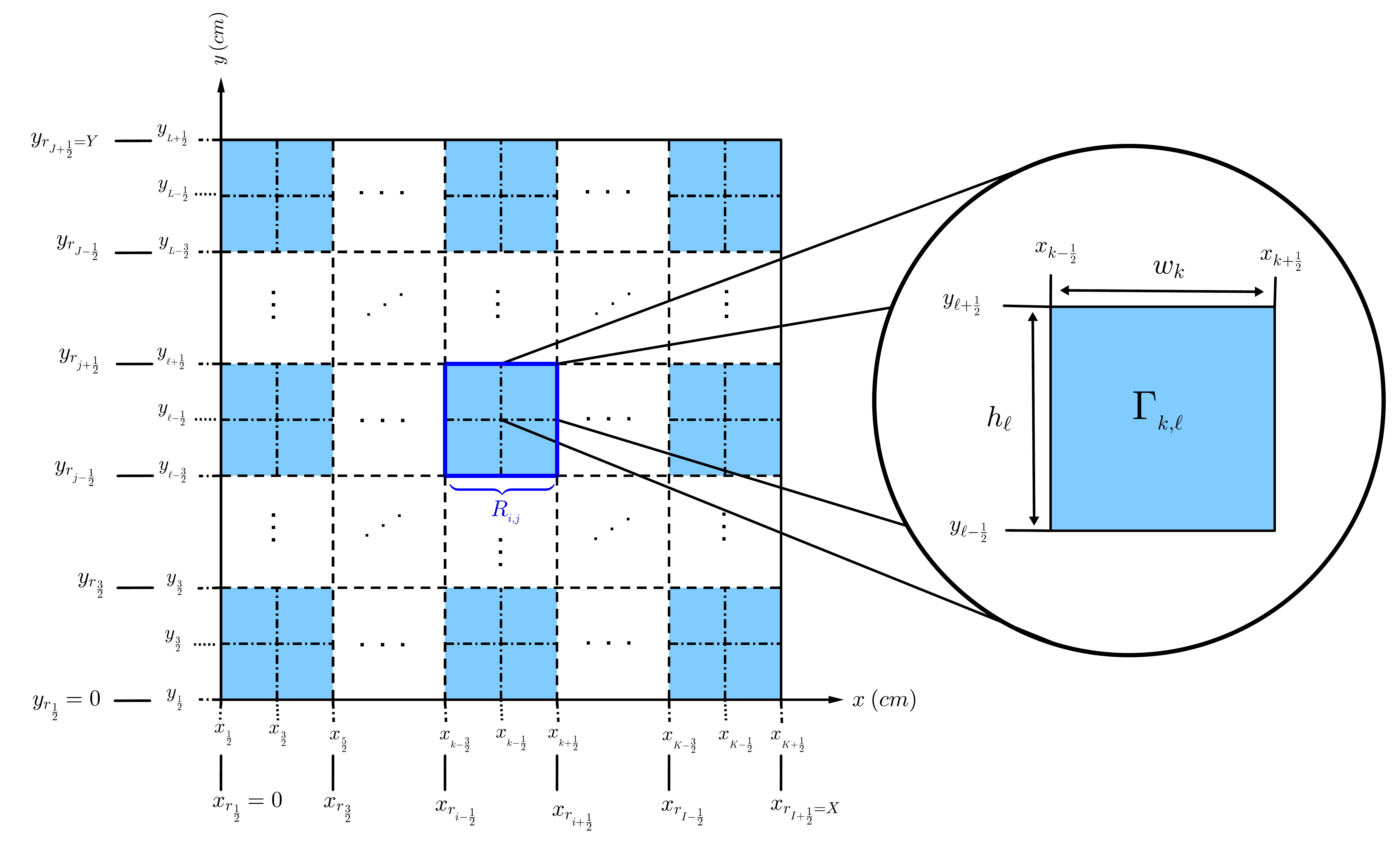}\vspace*{0cm}
 	\caption{Spatial discretization grid on the two-dimensional domain illustrated in \cref{f1}, considering $n_{x_{i}} = 2$ for $i=1:I$ and $n_{y_{j}} = 2$ for $j=1:J$.}
 	\label{f2}
 \end{figure}

To describe the \mr method, we first discretize the two-dimensional domain as depicted in \cref{f1}. Therefore, we discretize each region $R_{_{i,j}}$ in $n_{x_{i}}$ nodes in the $x$ coordinate direction and $n_{y_{j}}$ nodes in the $y$ direction, thus generating a rectangular spatial grid on the two-dimensional domain, wherein each discretization node is denoted by $\Gamma_{_{k,\ell}}$ of width $w_{_{k}}$ and height $h_{_{\ell}}$, with $k=1:K$ and $\ell = 1:L$. Figure \ref{f2} illustrates the discretization spatial grid on the two-dimensional domain, considering $n_{x_{i}} = 2$ for $i=1:I$ and $n_{y_{j}} = 2$ for $j=1:J$. The material parameters and sources are uniform inside nodes $\Gamma_{_{k,\ell}}$ and equal to the material parameters and sources of region $R_{_{i,j}}$ within which node $\Gamma_{_{k,\ell}}$ is located.

To proceed, we perform the following three steps: 
\begin{itemize}
 \item[(i)] transverse-integrate \cref{e2.5a} within node $\Gamma_{_{k,\ell}}$ over the $x$ and $y$ coordinate directions;
 \item[(ii)] define the group transverse-integrated adjoint angular fluxes \begin{subequations}
	\begin{equation}\label{e3.1a}
	\widetilde{\Psi}^{\dagger}_{_{m,g,\ell}}(x) = \frac{1}{h_{_{\ell}}}\int_{y_{\ell-\frac{1}{2}}}^{y_{\ell+\frac{1}{2}}}\Psi^{\dagger}_{_{g}}(x,y,\mu_{_{m}},\eta_{_{m}})dy
\end{equation}
and
\begin{equation}\label{e3.1b}
\widehat{\Psi}^{\dagger}_{_{m,g,k}}(y) = \frac{1}{w_{_{k}}}\int_{x_{k-\frac{1}{2}}}^{x_{k+\frac{1}{2}}}\Psi^{\dagger}_{_{g}}(x,y,\mu_{_{m}},\eta_{_{m}})dx;
\end{equation}	
\label[pluraleq]{e3.1}
\end{subequations}
\item[(iii)] approximate the terms corresponding to the transverse leakage terms for the multigroup transverse-integrated \SN transport equations by their node-edge averages. 
\end{itemize}

 By these steps, we obtain the energy multigroup adjoint transverse-integrated \SN constant nodal equations. That is,
\begin{subequations}
		\begin{equation}\label{e3.2a}
	\begin{aligned}
	&-\mu_{_{m}}\frac{d}{dx}{\widetilde{\Psi}}^{\dagger}_{_{m,g,\ell}}(x) - \frac{\eta_{_{m}}}{h_{_{\ell}}}\left[{\widehat{\Psi}}^{\dagger}_{_{m,g,k,\ell+\frac{1}{2}}}-{\widehat{\Psi}}^{\dagger}_{_{m,g,k,\ell-\frac{1}{2}}}\right] + \sigma_{t_{g,k,\ell}}{\widetilde{\Psi}}^{\dagger}_{_{m,g,\ell}}(x) \\& = \frac{1}{4}\sum_{g^{\prime} = 1}^{G}\sum_{n = 1}^{M}\left(\sigma_{s_{g \rightarrow g^{\prime},k,\ell}}+\chi_{_{g',k,\ell}}\,\nu\sigma_{f_{g,k,\ell}}\right){\widetilde{\Psi}}^{\dagger}_{_{n,g',\ell}}(x)\tau_{_{n}} + Q^{\dagger}_{_{g,k,\ell}},
	\end{aligned}  
	\end{equation}
	and
	\begin{equation}\label{e3.2b}\begin{aligned}
	&-\eta_{_{m}}\frac{d}{dy}{\widehat{\Psi}}^{\dagger}_{_{m,g,k}}(y) - \frac{\mu_{_{m}}}{w_{_{k}}}\left[ {\widetilde{\Psi}}^{\dagger}_{_{m,g,k+\frac{1}{2},\ell}}-{\widetilde{\Psi}}^{\dagger}_{_{m,g,k-\frac{1}{2},\ell}}\right] + \sigma_{t_{g,k,\ell}}{\widehat{\Psi}}^{\dagger}_{_{m,g,k}}(y)  \\& =\frac{1}{4}\sum_{g^{\prime} = 1}^{G}\sum_{n = 1}^{M}\left(\sigma_{s_{g \rightarrow g^{\prime},k,\ell}}+\chi_{_{g',k,\ell}}\,\nu\sigma_{f_{g,k,\ell}}\right){\widehat{\Psi}}^{\dagger}_{n,g',k}(y)\tau_{_{n}} + Q^{\dagger}_{_{g,k,\ell}},
	\end{aligned}
	\end{equation}
where $m = 1:M$, $g=1:G$ and $(x,y) \in \Gamma_{_{k,\ell}}$, with $k=1:K$ and $\ell=1:L$. Moreover, transverse-integrating \cref{e2.5c,e2.5d,e2.5e,e2.5f} we obtain the boundary equations  applied to \cref{e3.2a,e3.2b}.
	\label[pluraleq]{e3.2}
\end{subequations}

As the problem stated by \cref{e3.2} is linear, we write its solution as a superposition of its homogeneous and particular solutions. In other words,
\begin{subequations}
	\begin{equation}\label{e3.3a}
{\widetilde{\Psi}}^{\dagger}_{_{m,g,\ell}}(x) = \widetilde{\Psi}^{\dagger^{^{\mathcal{H}}}}_{_{m,g,\ell}}(x) + \widetilde{\Psi}^{\dagger^{^{\mathcal{P}}}}_{_{m,g,k,\ell}}
	\end{equation}
	and
	\begin{equation}\label{e3.3b}
{\widehat{\Psi}}^{\dagger}_{_{m,g,k}}(y) = {\widehat{\Psi}}^{\dagger^{^{\mathcal{H}}}}_{_{m,g,k}}(y) + {\widehat{\Psi}}^{\dagger^{^{\mathcal{P}}}}_{_{m,g,k,\ell}}
	\end{equation}
where the superscripts $\mathcal{H}$ and $\mathcal{P}$ denote the homogeneous and particular solutions respectively.

		\label[pluraleq]{e3.3}
\end{subequations}
\subsection{Homogeneous solutions to the multigroup adjoint transverse-integrated \SN constant nodal equations}\label{sec3.1}
Based on the literature \cite{MoraesPhy,Moraes20,Curbelo2021},
we write the homogeneous solution of \cref{e3.2a} within a node $\Gamma_{_{k,\ell}}$ as
\begin{equation}\label{e3.4}
{\widetilde{\Psi}}^{\dagger^{^{\mathcal{H}}}}_{_{m,g,\ell}}(x) = a_{_ {m,g}}(\vartheta)e^{-\frac{(x-\gamma)}{\vartheta}},\:\:\:\: \gamma =  \left \{
\begin{array}{cc}
x_{k-1/2},& \vartheta>0\\
x_{k+1/2},& \vartheta<0\\
\end{array}\right.,
\end{equation}
where the variable $\gamma$ indicates the application of the exponential shift procedure \cite{Jesusaaa,Cotta,Garcia} used to avoid numerical overflows due to finite computational arithmetic. Substituting \cref{e3.4} into \cref{e3.2a} with zero transverse leakage and zero adjoint source, we obtain the relation
\begin{equation}\label{e3.5}
\sum_{g^{\prime} = 1}^{G}\sum_{n=1}^{M}\left[\frac{\left(\sigma_{s_{g \rightarrow g^{\prime},k,\ell}}+\chi_{_{g',k,\ell}}\,\nu\sigma_{f_{g,k,\ell}}\right)\omega_{_{n}}}{4\mu_{_{m}}}- \frac{\delta_{_{m,n}}\delta_{_{g,g^{\prime}}}\sigma_{t_{g',k,\ell}}}{\mu_{_{m}}}\right]a_{_{n,g}}(\vartheta) = \frac{1}{\vartheta}a_{_{m,g}}(\vartheta),
\end{equation}
where $\delta$ stands for the kronecker delta. Now, fixing $m$ equal to 1 in \cref{e3.5}, varying $g$ from $1$ to $G$ and then repeating this process until $m \text{ is equal to }M$, we obtain an eigenvalue problem of order $MG$. Once the eigenvalue problem is solved, we write the homogeneous solution of \cref{e3.2a} as linear combination of the solutions proposed in \cref{e3.4} for each value of $\vartheta$. Thus,
\begin{equation}\label{e3.6}
{\widetilde{\Psi}}^{\dagger^{^{\mathcal{H}}}}_{_{m,g,\ell}}(x) = \sum_{l=1}^{MG}\alpha_{_{l}}\,a_{_{m,g}}(\vartheta_{_{l}})e^{-\frac{(x-\gamma)}{\vartheta_{_{l}}}},
\end{equation}
where $\alpha_{_{l}}$ are arbitrary constants. 

\subsection{General solution of the multigroup adjoint transverse-integrated \SN constant nodal equations}\label{sec3.2}
Having found the homogeneous solution in \cref{sec3.1}, we seek, at first, to obtain the particular solution of \cref{e3.2a}. Thus, we substitute ${\widetilde{\Psi}}^{\dagger^{^{\mathcal{P}}}}_{_{m,g,k,\ell}}$ into \cref{e3.2a}, to obtain the matrix equation for the particular solution whitin each discretization node
\begin{equation}\label{e3.9}
\boldsymbol{{\widetilde{\Psi}}}^{\dagger^{^{\mathcal{P}}}}_{k,\ell} = \boldsymbol{\Upsilon}_{_{k,\ell}}^{^{-1}}\left[\boldsymbol{Q}^{\dagger}_{_{k,\ell}} + \boldsymbol{I}_{\frac{\eta}{h_{_{\ell}}}}\left(\boldsymbol{{\Psi}}^{\dagger^{^{x}}}_{_{k,\ell+\frac{1}{2}}}- \boldsymbol{{\Psi}}^{\dagger^{^{x}}}_{_{k,\ell-\frac{1}{2}}}\right)\right],
\end{equation}
where $\boldsymbol{\Upsilon}_{_{k,\ell}}$ is a square matrix of order $MG$ composed of the material parameters, $\boldsymbol{I}_{\frac{\eta}{h_{_{\ell}}}}$ is a diagonal matrix of order $MG$ whose non-zero entries are $\frac{\eta}{h_{_{\ell}}}$ and $\boldsymbol{Q}^{\dagger}_{_{k,\ell}}$ is a $MG$-dimensional vector composed of the adjoint sources. Therefore, considering \cref{e3.3a,e3.9} we write the general solution of \cref{e3.2a} as
\begin{equation}\label{e3.10}
\boldsymbol{{\widetilde{\Psi}}}^{\dagger}_{\ell}(x)  = \boldsymbol{A}_{k,\ell}(x)\boldsymbol{\alpha}_{k,\ell} + \boldsymbol{\Upsilon}_{_{k,\ell}}^{^{-1}}\left[\boldsymbol{Q}^{\dagger}_{_{k,\ell}} + \boldsymbol{I}_{\frac{\eta}{h_{_{\ell}}}}\left(\boldsymbol{{\Psi}}^{\dagger^{^{x}}}_{_{k,\ell+\frac{1}{2}}}- \boldsymbol{{\Psi}}^{\dagger^{^{x}}}_{_{k,\ell-\frac{1}{2}}}\right)\right].
\end{equation}

To proceed, one follows similar steps to determine the general solution of \cref{e3.2b}.

\subsection{The \mr discretized equations}\label{sec3.3}
To determine the adjoint node-edge average angular fluxes cf. \cref{e3.10} it is necessary to calculate the arbitrary constants. Therefore, let us initially define the $MG$-dimensional column vectors
\begin{subequations}
\begin{equation}\label{e3.11a}
	\boldsymbol{\widetilde{\Psi}}^{\dagger^{^{(in/out)}}}_{_{\ell}} = \left[ \boldsymbol{\widetilde{\Psi}}^{\dagger}_{_{1q,k \lmp \frac{1}{2},\ell}},\boldsymbol{\widetilde{\Psi}}^{\dagger}_{_{2q,k \lpm \frac{1}{2},\ell}},\boldsymbol{\widetilde{\Psi}}^{\dagger}_{_{3q,k \lpm \frac{1}{2},\ell}},\boldsymbol{\widetilde{\Psi}}^{\dagger}_{_{4q,k \lmp \frac{1}{2},\ell}} \right]^{^{\text{T}}}
\end{equation}
and
\begin{equation}\label{e3.11b}
\boldsymbol{\widehat{\Psi}}^{\dagger^{^{(in/out)}}}_{_{k}}= \left[ \boldsymbol{\widehat{\Psi}}^{\dagger}_{_{1q,k,\ell\lmp \frac{1}{2}}},\boldsymbol{\widehat{\Psi}}^{\dagger}_{_{2q,k,\ell\lmp \frac{1}{2}}},\boldsymbol{\widehat{\Psi}}^{\dagger}_{_{3q,k,\ell\lpm \frac{1}{2}}},\boldsymbol{\widehat{\Psi}}^{\dagger}_{_{4q,k,\ell\lpm \frac{1}{2}}} \right]^{^{\text{T}}},
\end{equation}
where superscripts $(in)$ and $(out)$ indicate that the adjoint node-edge average angular fluxes are evaluated in the incoming and outgoing directions of node $\Gamma_{_{k,\ell}}$ respectively. In addition, the subscript $u\,q$, with $u = \left\lbrace1,2,3,4 \right\rbrace $, indicates which set of discrete directions are being considered in the adjoint node-edge average angular fluxes, viz \cref{f3}. Figure \ref{f3} displays a schematic illustration of the adjoint node-edge average angular fluxes in node $\Gamma_{_{k,\ell}}$, 
 \begin{figure}
	\centering
	\includegraphics[scale=0.27]{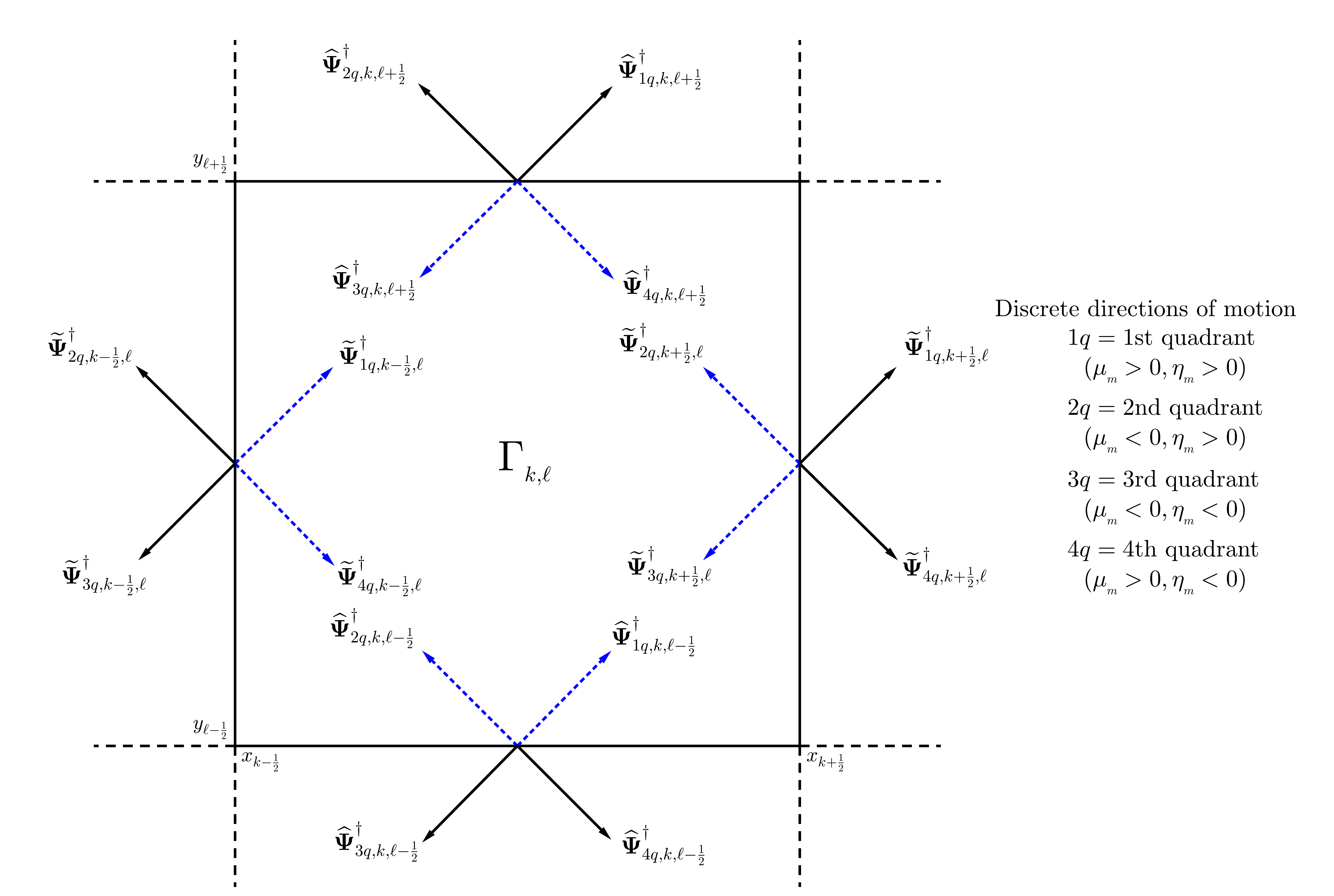}\vspace*{0cm}
	\caption{Schematic illustration of the adjoint node-edge average angular fluxes in the incoming and outgoing directions of node $\Gamma_{_{k,\ell}}$.}
	\label{f3}
\end{figure}where the dashed arrows represent the adjoint node-edge fluxes in the incoming directions and the solid arrows represent the adjoint node-edge fluxes in the outgoing directions. In addition, \cref{f3} also illustrates that the discrete directions are divided into four quadrants which are ordered in a counterclockwise fashion. At this point, we remark that \cref{e3.11} have the same matrix definitions as \cref{e3.10}. Therefore, for each fixed discrete direction $m$, we vary $g$ from $1$ to $G$. 
		\label[pluraleq]{e3.11}
\end{subequations}

To proceed, we make use of \cref{e3.10} and \cref{e3.11}, to write
\begin{subequations}
	\begin{equation}\label{e3.12a}
	\boldsymbol{{\widetilde{\Psi}}}^{\dagger^{^{^{(out)}}}}_{_{\ell}} = \boldsymbol{\mathcal{A}}_{_{k,\ell}}\boldsymbol{\alpha}_{_{k,\ell}}  + \boldsymbol{\Upsilon}_{_{k,\ell}}^{^{-1}}\left[\boldsymbol{Q}^{\dagger}_{_{k,\ell}} + \boldsymbol{I}_{\frac{\eta}{h_{_{\ell}}}}\boldsymbol{U}\left(\boldsymbol{{\widehat{\Psi}}}^{\dagger^{^{(in)}}}_{_{k}}- \boldsymbol{{\widehat{\Psi}}}^{\dagger^{^{(out)}}}_{_{k}}\right)\right],
	\end{equation}
	where 
	\begin{equation}\label{e3.12b}
	 \boldsymbol{\mathcal{A}}_{_{k,\ell}} = \boldsymbol{E}_{1}\boldsymbol{A}_{_{k,\ell}}(x_{k+\frac{1}{2}}) + \boldsymbol{E}_{2}\boldsymbol{A}_{_{k,\ell}}(x_{k-\frac{1}{2}})+\boldsymbol{E}_{3}\boldsymbol{A}_{_{k,\ell}}(x_{k-\frac{1}{2}})+\boldsymbol{E}_{4}\boldsymbol{A}_{_{k,\ell}}(x_{k+\frac{1}{2}})
	\end{equation}
	and
	\begin{equation}\label{e3.12c}
	\boldsymbol{U} = -\boldsymbol{E}_{1}-\boldsymbol{E}_{2}+\boldsymbol{E}_{3}+\boldsymbol{E}_{4}.
	\end{equation}
	In \cref{e3.12b,e3.12c} $\boldsymbol{E}_{n}$ are square matrices of order $MG$ whose entries are
\begin{eqnarray}
E_{a,b}= \left \{
\begin{array}{cc}
\boldsymbol{I}_{_{\frac{MG}{4}}},& a = b = n \\
\boldsymbol{O}_{_{\frac{MG}{4}}},& \text{otherwise} \\
\end{array}
\right., \text{for  } a,b = 1:4, \nonumber
\end{eqnarray}
where $\boldsymbol{I}_{_{\frac{MG}{4}}}$ is the identity matrix of order $\frac{MG}{4}$ and $\boldsymbol{O}_{_{\frac{MG}{4}}}$ is the zero-matrix of order $\frac{MG}{4}$. The arbitrary constants can be determined directly by \cref{e3.12a} as 
\begin{equation}\label{e3.12d}
 \boldsymbol{\alpha}_{_{k,\ell}} = \boldsymbol{\mathcal{A}}_{_{k,\ell}}^{^{-1}}\left\lbrace 	\boldsymbol{{\widetilde{\Psi}}}^{\dagger^{^{^{(out)}}}}_{_{\ell}} -  \boldsymbol{\Upsilon}_{_{k,\ell}}^{^{-1}}\left[\boldsymbol{Q}^{\dagger}_{_{k,\ell}} + \boldsymbol{I}_{\frac{\eta}{h_{_{\ell}}}}\boldsymbol{U}\left(\boldsymbol{{\widehat{\Psi}}}^{\dagger^{^{(in)}}}_{_{k}}- \boldsymbol{{\widehat{\Psi}}}^{\dagger^{^{(out)}}}_{_{k}}\right)\right]    \right\rbrace.
\end{equation}
\label[pluraleq]{e3.12}
\end{subequations}

By substituting \cref{e3.12d} into \cref{e3.10} we obtain
	\begin{equation}\label{e3.13}
\adjustbox{scale=0.965}{$\boldsymbol{{\widetilde{\Psi}}}^{\dagger}_{_{\ell}}(x) = \boldsymbol{A}_{_{k,\ell}}(x)\boldsymbol{\mathcal{A}}_{_{k,\ell}}^{^{-1}}\boldsymbol{{\widetilde{\Psi}}}^{\dagger^{^{^{(out)}}}}_{_{\ell}} + \left(\boldsymbol{I} - \boldsymbol{A}_{_{k,\ell}}(x)\boldsymbol{\mathcal{A}}_{_{k,\ell}}^{^{-1}} \right) \boldsymbol{\Upsilon}_{_{k,\ell}}^{^{-1}}\left[\boldsymbol{Q}^{\dagger}_{_{k,\ell}} + \boldsymbol{I}_{\frac{\eta}{h_{_{\ell}}}}\boldsymbol{U}\left(\boldsymbol{{\widehat{\Psi}}}^{\dagger^{^{(in)}}}_{_{k}}- \boldsymbol{{\widehat{\Psi}}}^{\dagger^{^{(out)}}}_{_{k}}\right)\right],$}
	\end{equation}
where $\boldsymbol{I}$ is the identity matrix of order $MG$. By using \cref{e3.13} and \cref{e3.11a}, we obtain
\begin{subequations}
	\begin{equation}\label{e3.14a}
	\boldsymbol{{\widetilde{\Psi}}}^{\dagger^{^{^{(in)}}}}_{_{\ell}}= \boldsymbol{\Lambda}_{_{k,\ell}}\boldsymbol{{\widetilde{\Psi}}}^{\dagger^{^{^{(out)}}}}_{_{\ell}} + \left(\boldsymbol{I} - \boldsymbol{\Lambda}_{_{k,\ell}} \right) \boldsymbol{\Upsilon}_{_{k,\ell}}^{^{-1}}\left[\boldsymbol{Q}^{\dagger}_{_{k,\ell}} + \boldsymbol{I}_{\frac{\eta}{h_{_{\ell}}}}\boldsymbol{U}\left(\boldsymbol{{\widehat{\Psi}}}^{\dagger^{^{(in)}}}_{_{k}}- \boldsymbol{{\widehat{\Psi}}}^{\dagger^{^{(out)}}}_{_{k}}\right)\right],
	\end{equation}
	where 
	\begin{equation}\label{e3.14b}
	\boldsymbol{\Lambda}_{_{k,\ell}} = \left\lbrace  \boldsymbol{E}_{1}\boldsymbol{A}_{_{k,\ell}}(x_{k-\frac{1}{2}}) + \boldsymbol{E}_{2}\boldsymbol{A}_{_{k,\ell}}(x_{k+\frac{1}{2}})+\boldsymbol{E}_{3}\boldsymbol{A}_{_{k,\ell}}(x_{k+\frac{1}{2}})+\boldsymbol{E}_{4}\boldsymbol{A}_{_{k,\ell}}(x_{k-\frac{1}{2}})\right\rbrace \boldsymbol{\mathcal{A}}_{_{k,\ell}}^{^{-1}}.
	\end{equation}
	\label[pluraleq]{e3.14}
\end{subequations}

Equation (\ref{e3.14a}) together with the corresponding equation in the $y$ coordinate direction, i.e.,
\begin{subequations}
	\begin{equation}\label{e3.15a}
	\boldsymbol{{\widehat{\Psi}}}^{\dagger^{^{^{(in)}}}}_{_{k}}= \boldsymbol{\Lambda}_{_{k,\ell}}'\boldsymbol{{\widehat{\Psi}}}^{\dagger^{^{^{(out)}}}}_{_{k}} + \left(\boldsymbol{I} - \boldsymbol{\Lambda}_{_{k,\ell}}' \right) \boldsymbol{\Upsilon}_{_{k,\ell}}^{^{-1}}\left[\boldsymbol{Q}^{\dagger}_{_{k,\ell}} + \boldsymbol{I}_{\frac{\mu}{w_{k}}}\boldsymbol{U}'\left(\boldsymbol{{\widetilde{\Psi}}}^{\dagger^{^{^{(in)}}}}_{_{\ell}}- \boldsymbol{{\widetilde{\Psi}}}^{\dagger^{^{^{(out)}}}}_{_{\ell}}\right)\right],
\end{equation}
are the discretized equations of the \mr method. In \cref{e3.15a} we have defined 
\begin{equation}\label{e3.15b}
\boldsymbol{\Lambda}_{_{k,\ell}}' =  \left\lbrace  \boldsymbol{E}_{1}\boldsymbol{A}_{_{k,\ell}}'(y_{\ell-\frac{1}{2}}) + \boldsymbol{E}_{2}\boldsymbol{A}_{_{k,\ell}}'(y_{\ell-\frac{1}{2}})+\boldsymbol{E}_{3}\boldsymbol{A}_{_{k,\ell}}'(y_{\ell+\frac{1}{2}})+\boldsymbol{E}_{4}\boldsymbol{A}_{_{k,\ell}}'(y_{\ell+\frac{1}{2}})\right\rbrace \boldsymbol{\mathcal{A}}_{_{k,\ell}}'^{^{-1}}
\end{equation}
and 
	\begin{equation}\label{e3.15c}
\boldsymbol{U}' = -\boldsymbol{E}_{1}+\boldsymbol{E}_{2}+\boldsymbol{E}_{3}-\boldsymbol{E}_{4},
\end{equation}
where $\boldsymbol{A}_{_{k,\ell}}'(y)$ is a square matrix of order $MG$ that composes the general solution of \cref{e3.2b} in an analogous way as $\boldsymbol{A}_{_{k,\ell}}(x)$ in \cref{e3.10}. Moreover, matrix $\boldsymbol{\mathcal{A}}_{_{k,\ell}}'$ in \cref{e3.15b} has the form
\begin{equation}\label{e3.15d}
	 \boldsymbol{\mathcal{A}}_{_{k,\ell}}' = \boldsymbol{E}_{1}\boldsymbol{A}_{_{k,\ell}}'(y_{\ell+\frac{1}{2}}) + \boldsymbol{E}_{2}\boldsymbol{A}_{_{k,\ell}}'(y_{\ell+\frac{1}{2}})+\boldsymbol{E}_{3}\boldsymbol{A}_{_{k,\ell}}'(y_{\ell-\frac{1}{2}})+\boldsymbol{E}_{4}\boldsymbol{A}_{_{k,\ell}}'(y_{\ell-\frac{1}{2}}).
\end{equation}

The discretized equations of the \mr method are used to implement an iterative numerical scheme to converge results for the adjoint node-edge average angular fluxes. In this work we use the adjoint partial one-node block inversion (partial NBI$^{\dagger}$) iterative scheme that uses the boundary conditions and the most recent estimates available for the adjoint node-edge average angular fluxes in the outgoing directions to evaluate the incoming fluxes that constitute the exiting fluxes for the adjacent nodes in the directions of the transport sweeps. The partial NBI$^{\dagger}$ iterative scheme can be found in more details in the literature \cite{Curbelo2021}.
\label[pluraleq]{e3.15}
\end{subequations}
 
 Analyzing the \mr discretized equations we can notice that all sweeping matrices do not change with the variation of the adjoint source term. In other words, one motivation for choosing the \mr method for this type of problem is due to the fact that the sweeping matrices built for generating the \mr discretized equations need to be calculated just once, regardless of how many times we need to solve the adjoint equations in order to build the importance matrix. In each process of solving the adjoint equations, the only changing variable is the adjoint source term; therefore one just needs to recalculate $\boldsymbol{Q}^{\dagger}_{_{k,\ell}}$, cf. \cref{e3.14a,e3.15a}. Therefore, as opposed to conventional methods, such as the DD method, that a full run of the computer code is required due to the change of the adjoint source term, as with the \mr method we run the code starting at the implementation of the partial NBI$^{\dagger}$ scheme using the appropriate matrix $\boldsymbol{Q}^{\dagger}_{_{k,\ell}}$.

\section{Numerical results}\label{sec4}
\setcounter{equation}{0}
 
 We present in this section numerical experiments to two model problems in order to: (i) analyze characteristics of the importance function; (ii) illustrate the influence of the auxiliary equations over the results; and (iii) check the accuracy of the present methodology. For these numerical experiments we consider that the average number of neutrons released in fission is $\nu =3$ and the average energy released in one fission event is $\epsilon = 200\,MeV$. Moreover, the stopping criterion adopted in the partial NBI$^{\dagger}$ requires that the relative deviations between two consecutive estimates of adjoint node-edge average scalar fluxes be no greater than $1\times 10^{-5}$.

\subsection{Model problem I}\label{sec4.1}

\begin{figure}
	\centering
	\includegraphics[scale=0.25]{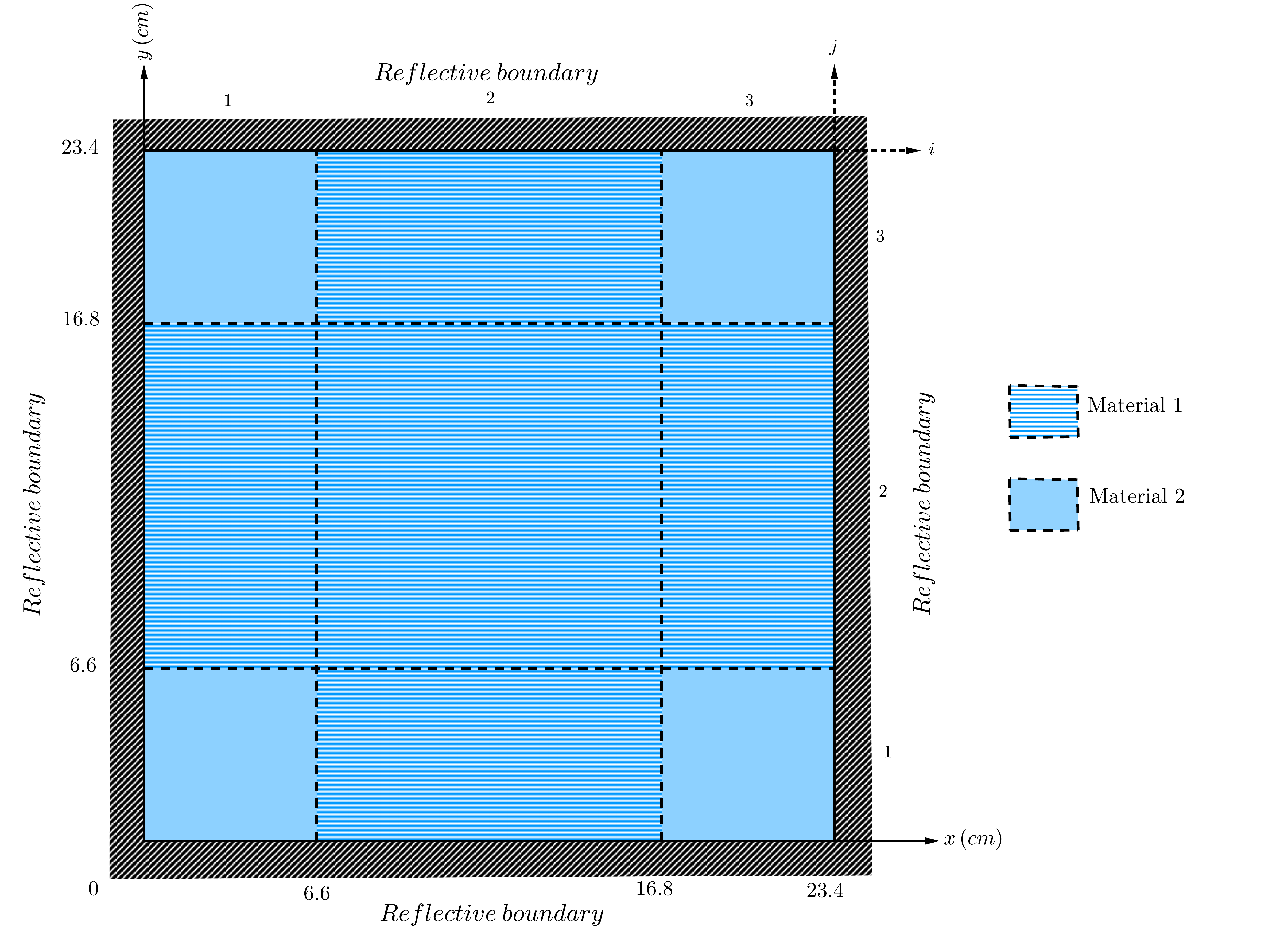}\vspace*{0cm}
	\caption{Model problem I.}
	\label{f4}
\end{figure}

For the first model problem we consider a square domain composed of 9 fuel regions and two different materials, as illustrated in \cref{f4}. In addition, \cref{t1} lists the material parameters considered in model problem I. The discretization spatial grid is defined such that $n_{x_{i}} = n_{y_{i}} = 44$ for $i = 1 \text{ or }3$ and $n_{x_{2}} = n_{y_{2}} = 68$. Thus, a uniform spatial grid composed of $156\times156$ nodes was set upon the domain displayed in \cref{f4}. Moreover, we used to model the adjoint problems the LQ$_{\text{8}}$ formulation and the multigroup formulation with $G = 4$ energy groups. With this configuration, the multiplication factor of the system is $k_{eff} = 0.64581$, which ensures the subcritical condition. Thus, one must add uniform source of neutrons to this subcritical system in order to drive it to a prescribed steady-state power density level. Therefore, we follow the offered methodology to determine these neutron source distributions, considering that the power generated by all fuel regions per unit length of core height is $P_{_{total}} = 10\,MWcm^{-1}$. 
 				{\renewcommand{\arraystretch}{1.3}
\begin{table}
	\centering
	\caption{Material parameters of model problem I.}\label{t1}
	\small
	\begin{threeparttable}
		\adjustbox{max width=\textwidth}{%
		 \begin{tabular}{cc|c|c|cccc|c}
			\toprule[0,5mm]
			\multicolumn{2}{c|}{\multirow{2}[0]{*}{Material parameters}} & \multirow{2}[0]{*}{$\sigma_{t_{g}} (cm^{-1})$} & \multirow{2}[0]{*}{$\sigma_{f_{g}} (cm^{-1})$} & \multicolumn{4}{c|}{$\sigma_{s_{g\rightarrow g^{\prime}}} (cm^{-1})$} & \multirow{2}[0]{*}{$\chi_{g}$} \\
			\cline{5-8}    \multicolumn{2}{c|}{} &       &       & $g^{\prime} = 1$ & $g^{\prime} = 2$ & $g^{\prime} = 3$ & $g^{\prime} =4$ &  \\
			\hline
			\multirow{4}[2]{*}{Material 1} & $g = 1$ & 2.37640E-01\tnote{a} & 2.74720E-03 & 1.66870E-01 & 6.67810E-02 & 2.95390E-04 & 9.80000E-08 & 1 \\
			& $g = 2$ & 5.46200E-01 & 2.41330E-04 & 0.00000E+00 & 4.84360E-01 & 5.92300E-02 & 1.93200E-05 & 0 \\
			& $g = 3$ & 9.16560E-01 & 3.22790E-03 & 0.00000E+00 & 0.00000E+00 & 8.36010E-01 & 6.06480E-02 & 0 \\
			& $g = 4$ & 1.18980E+00 & 3.30453E-02 & 0.00000E+00 & 0.00000E+00 & 0.00000E+00 & 1.11550E+00 & 0 \\
			\hline
			\multirow{4}[2]{*}{Material 2} & $g = 1$ & 2.16160E-01 & 1.94130E-04 & 1.39920E-01 & 5.38490E-02 & 2.44710E-04 & 8.07000E-08 & 1 \\
			& $g = 2$ & 4.56080E-01 & 2.04570E-04 & 0.00000E+00 & 3.87780E-01 & 4.79060E-02 & 1.57160E-05 & 0 \\
			& $g = 3$ & 9.58540E-01 & 2.12790E-04 & 0.00000E+00 & 0.00000E+00 & 6.53650E-01 & 3.11650E-02 & 0 \\
			& $g = 4$ & 2.49310E+00 & 3.40533E-02 & 0.00000E+00 & 0.00000E+00 & 0.00000E+00 & 1.11010E+00 & 0 \\
			\bottomrule[0,5mm]
		\end{tabular}%
	}
		\normalsize
		\begin{tablenotes}[flushleft]
			\tiny
			\item[a] Read as $2.3764\times10^{-01}$.
		\end{tablenotes}
	\end{threeparttable}
\end{table}
}

\begin{figure}
	\begin{subfigure}{.5\textwidth}
		\centering
	\includegraphics[scale=0.5]{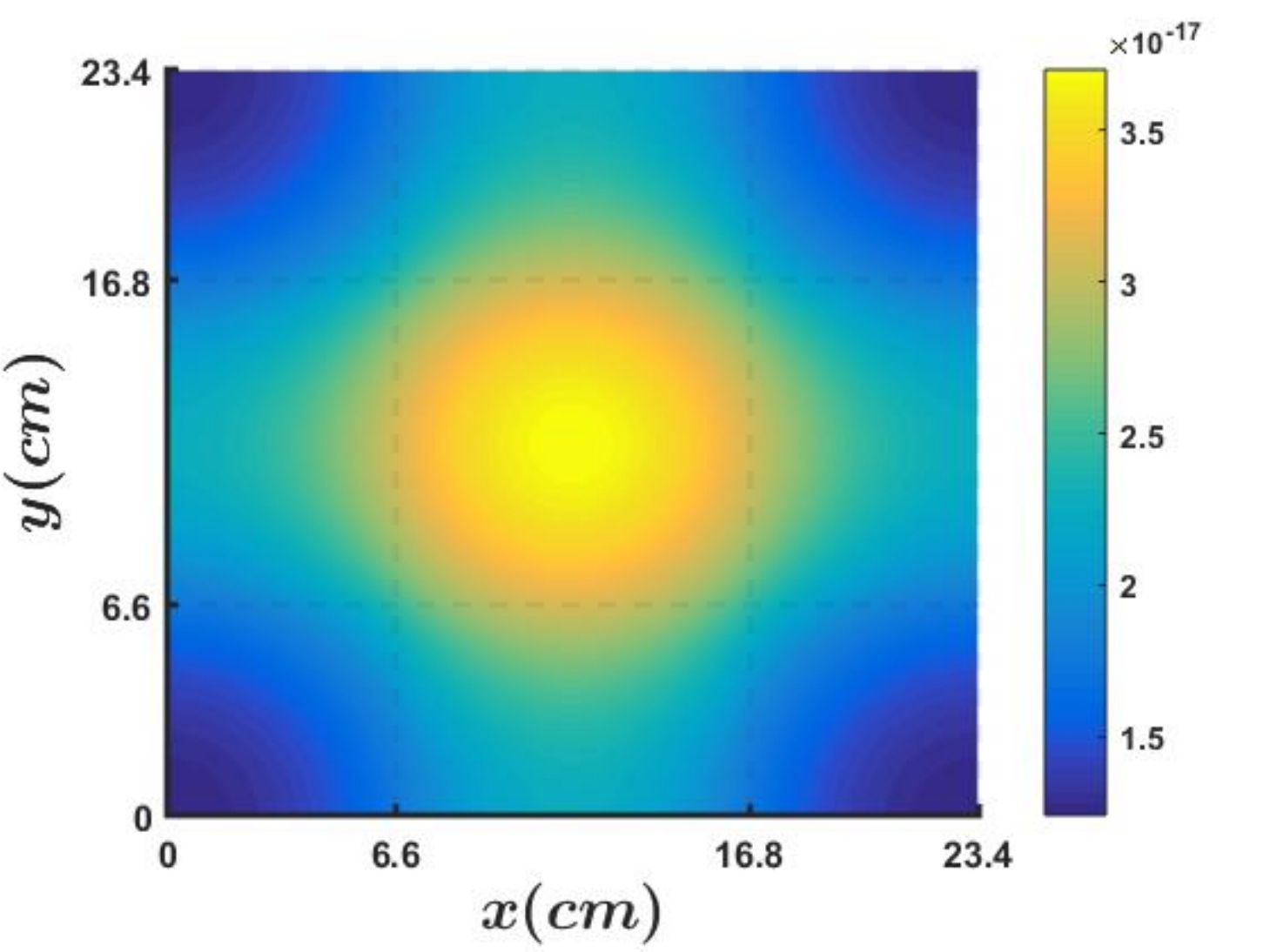}	
	\caption{$g = 1$}
	\end{subfigure}
	\begin{subfigure}{.5\textwidth}
		\centering
	\includegraphics[scale=0.5]{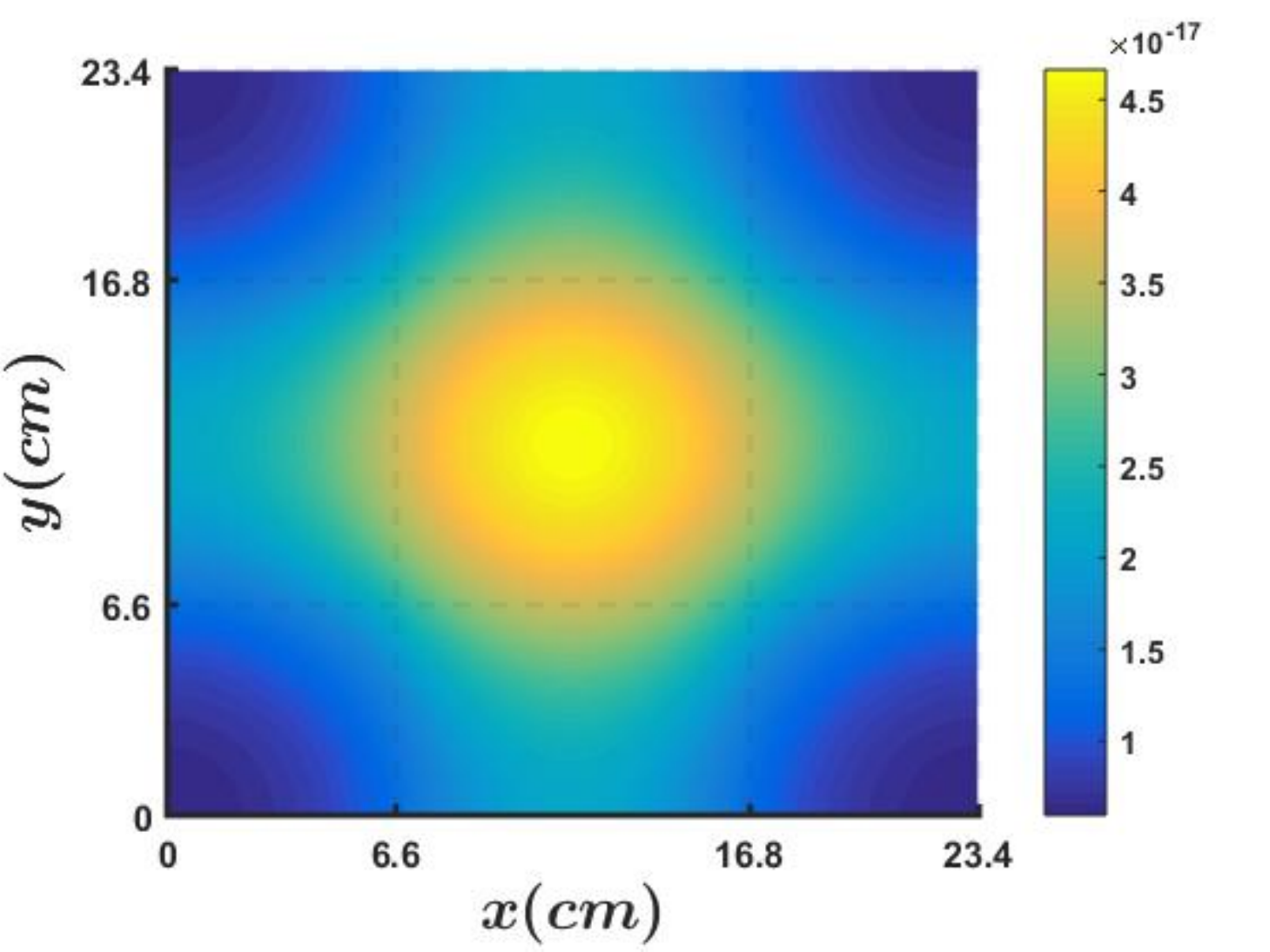}
		\caption{$g = 2$}
	\end{subfigure}
\\
	\begin{subfigure}{.5\textwidth}
	\centering
			\includegraphics[scale=0.5]{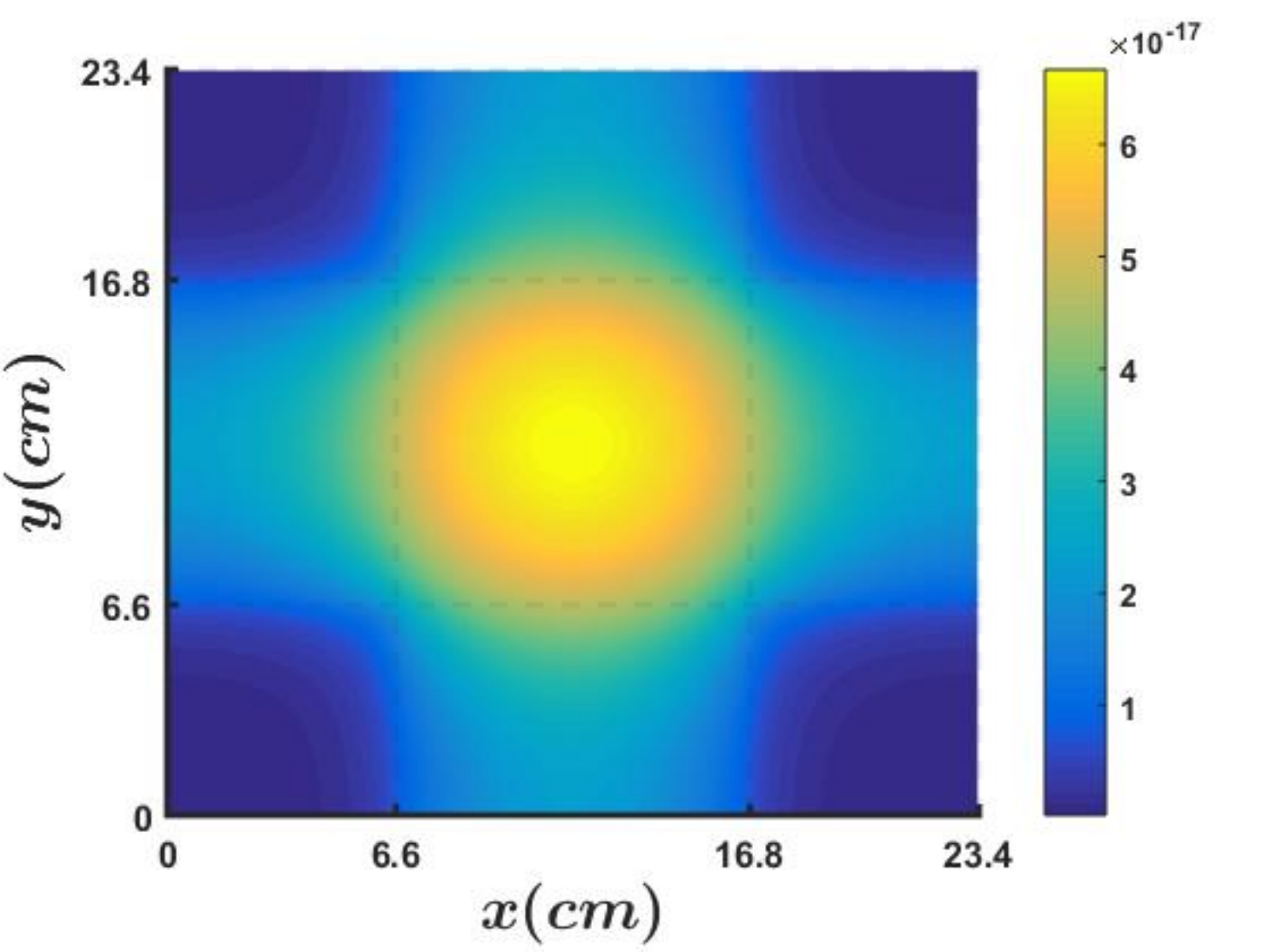}	
				\caption{$g = 3$}
\end{subfigure}
\begin{subfigure}{.5\textwidth}
	\centering
			\includegraphics[scale=0.5]{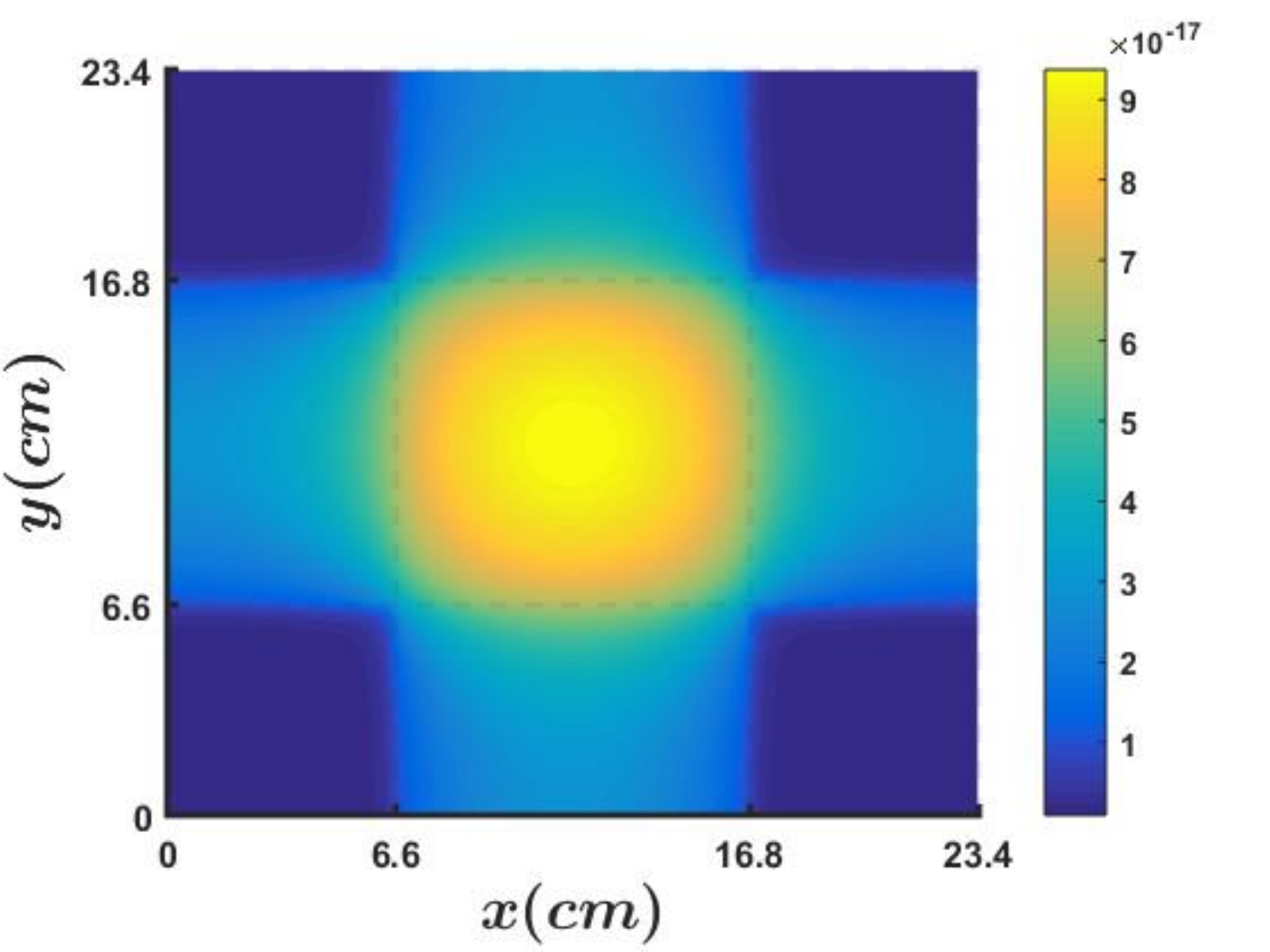}	
				\caption{$g = 4$}
\end{subfigure}
\caption{Node-average adjoint scalar fluxes considering the adjoint source term with $i^{\star},j^{\star} = 2$ in model problem I.}
\label{f5}
\end{figure}

We first built the importance matrix associated with this problem by solving \cref{2.5} 9 times using sequentially the adjoint source terms according to \cref{e2.7}. In each of these 9 problems, the adjoint source term changes, correlating the solution of \cref{2.5} with the power generated by one specific fuel region of the system. Figure \ref{f5} displays the group node-average adjoint 
scalar fluxes considering $i^{\star},j^{\star} = 2$ in \cref{e2.7}, as a function of the node midpoints.

Although in \cref{f5} we focus our attention in region $R_{_{2,2}}$, we can infer some general characteristics of the neutron importance with respect to generation of power. We see in \cref{f5} that all neutrons have importance to the generation of power. This characteristic can be explained by the fact that neutrons do eventually migrate to region $R_{_{2,2}}$ and induce fission.

Further, in \cref{f5} we can also note that neutrons migrating in regions composed of material 1 have higher importance to the generation of power than those migrating in regions composed of material 2 (corner regions). This fact is explained not only by the wider interfaces of these regions (material 1) with region $R_{_{2,2}}$, but also by the smaller probability that regions composed of material 1 have to undergo radiative capture events ($\sigma_{\gamma}$), viz \cref{t2}. As we see in \cref{t2} the radiative capture macroscopic cross sections of material 2 are larger than those of material 1. This means that neutrons migrating in regions composed of material 2 are subject to higher probabilities of absorption (radiative capture). Once neutrons are absorbed they will not contribute to the generation of power, hence having zero importance.
	{\renewcommand{\arraystretch}{1.3}
	\begin{table}[]{}
		\centering
		\caption{Radiative capture macroscopic cross sections of model problem I.}
		\small
		\begin{threeparttable}
			\begin{tabular}{c|cccc}
				\toprule[0.5mm]
				\multirow{2}[1]{*}{Material} & \multicolumn{4}{c}{$\sigma_{\gamma_{g}}$\tnote{b} $(cm^{-1})$} \\
				\cline{2-5}          & $g = 1$ & $g =2$ & $g = 3$ & $g = 4$ \\
				\hline
			1     & 9.46312E-04\tnote{a} & 2.34935E-03 & 1.66741E-02 & 4.12547E-02 \\
			\hline
			2     & 2.19521E-02 & 2.01737E-02 & 2.73512E-01 & 1.34895E+00 \\
				\bottomrule[0.5mm]
			\end{tabular}%
			\begin{tablenotes}[flushleft]
				\tiny
				\item[a] Read as 9.46312$\times10^{-04}$.
				\item[b] $\sigma_{\gamma_{g}} = \sigma_{t_{g}} - \sum\limits_{g^{\prime}=1}^{4}\sigma_{s_{g\rightarrow g'}} - \sigma_{f_{g}}$.
			\end{tablenotes}
			\normalsize
		\end{threeparttable}
		\label{t2}%
\end{table}}

Figure \ref{f5} also shows that neutrons migrating with different energy will contribute differently to the generation of power. As we see, neutrons migrating with lower kinetic energy ($g=4$) have more importance for the power in region $R_{2,2}$ than the others.

\subsubsection{Numerical experiment I}
For the first numerical experiment we consider that the total power density generated has to be $P_{_{total}} = 10\,MWcm^{-1}$. Since the importance matrix, as presented in \cref{e2.9b}, has already been calculated for model problem I, we can correlate the power of the whole system with the interior neutron sources by summing all rows of the importance matrix. As each row of the importance matrix is associated with the power generated by one specific fuel region, viz \cref{e2.8a,e2.9a}, the sum of all rows of this importance matrix is correlated with the power generated by the whole system ($P_{_{total}}$).  \cref{t3} displays the linear system of one equation and 36 unknwons built for the first numerical experiment and the entries of the row importance matrix of order $1\times 36$.
{\renewcommand{\arraystretch}{1.4}
\begin{table}[htbp]{}
	\centering
	\caption{The linear system built for the first numerical experiment.}
	\small
	\begin{threeparttable}
\begin{tabular}{!{\vrule width 1pt}c!{\vrule width 1pt} c!{\vrule width 1pt}c:c:c:c:c:c!{\vrule width 1pt} c!{\vrule width 1pt} c!{\vrule width 1pt}}
\multirow{12}{*}{$                 P_{_{total}}
$\tnote{a} } & \multirow{12}{*}{=} & 1,1,1\tnote{b}  & 2,1,1  & 3,1,1  & 4,1,1  & 1,2,1  & 2,2,1  & \multirow{12}{*}{} & \multirow{12}{*}{$Q_{g,i,j}$}\\
 &  & 2.15721E-15\tnote{c}\,\:  & 1.48987E-15  & 4.78797E-16  & 4.42264E-16  & 4.86619E-15  & 4.86789E-15  &  & \\
\cline{3-8} \cline{4-8} \cline{5-8} \cline{6-8} \cline{7-8} \cline{8-8} 
 &  & 3,2,1  & 4,2,1  & 1,3,1  & 2,3,1  & 3,3,1  & 4,3,1  &  & \tabularnewline
 &  & 5.74992E-15  & 8.02857E-15  & 2.15721E-15  & 1.48987E-15  & 4.78797E-16  & 4.42264E-16  &  & \tabularnewline
\cline{3-8} \cline{4-8} \cline{5-8} \cline{6-8} \cline{7-8} \cline{8-8} 
 &  & 1,1,2  & 2,1,2  & 3,1,2  & 4,1,2  & 1,2,2  & 2,2,2  &  & \tabularnewline
 &  & 4.86618E-15  & 4.86787E-15  & 5.7499E-15  & 8.02855E-15  & 9.73237E-15  & 9.73577E-15  &  & \tabularnewline
\cline{3-8} \cline{4-8} \cline{5-8} \cline{6-8} \cline{7-8} \cline{8-8} 
 &  & 3,2,2  & 4,2,2  & 1,3,2  & 2,3,2  & 3,3,2  & 4,3,2  &  & \tabularnewline
 &  & 1.00643E-14  & 1.10083E-14  & 3.1148E-15  & 2.3744E-15  & 5.34498E-15  & 5.31015E-15  &  & \tabularnewline
\cline{3-8} \cline{4-8} \cline{5-8} \cline{6-8} \cline{7-8} \cline{8-8} 
 &  & 1,1,3  & 2,1,3  & 3,1,3  & 4,1,3  & 1,2,3  & 2,2,3  &  & \tabularnewline
 &  & 2.15721E-15  & 1.48987E-15  & 4.78797E-16  & 4.42264E-16 & 4.86619E-15  & 4.86789E-15  &  & \tabularnewline
\cline{3-8} \cline{4-8} \cline{5-8} \cline{6-8} \cline{7-8} \cline{8-8} 
 &  & 3,2,3  & 4,2,3  & 1,3,3  & 2,3,3  & 3,3,3  & 4,3,3  &  & \tabularnewline
 &  & 5.74992E-15  & 8.02857E-15  & 2.15721E-15  & 1.48987E-15  & 4.78797E-16  & 4.42264E-16  &  & \tabularnewline
\end{tabular}
		\begin{tablenotes}[flushleft]
			\tiny
			\item[a] Row of the importance matrix correlated with the power generated by the whole system.
			\item[b] Element $\left(a_{_{1,1}}\right)$ of the row importance matrix that multiplies $Q_{1,1,1}$. The element index $1,z$ is defined such that $z = g + 4\left[ (i-1) + 3(j-1)\right] $.
			\item[c] Read as 2.15721$\times10^{-15}$.
			\item[-] Linear system of order $1\times 36$.
		\end{tablenotes}
	\end{threeparttable}
	\label{t3}%
\end{table}

The linear system presented in \cref{t3} does not possess a unique solution. Therefore, in order to obtain a unique solution for this problem we need to provide 35 auxiliary equations. In this numerical experiment we consider the neutron source distribution of the form
\begin{equation}\label{e4.1}
Q_{g,i,j} = \left \{
\begin{array}{cl}
\mathcal{Q},& \text{if}\:\: (g,i,j) = (2,2,2)\\
0.5\times \mathcal{Q},&  \text{if}\:\: (g,i,j) = (3,2,2)\\
0,& \text{otherwise}
\end{array}
\right..
\end{equation}
For the auxiliary equations presented in \cref{e4.1}, we obtain the unique solution of the linear system displayed in \cref{t3}: i.e., $\mathcal{Q} = 6.23748\times 10^{+14}$ neutrons $cm^{-3}s^{-1}$. This means that the neutron source distribution determined by the offered methodology is, in this case, composed of $Q_{2,2,2} =$ 6.23748$\times 10^{+14}$ neutrons $cm^{-3}s^{-1}$, $Q_{3,2,2} =$ 3.11874$\times 10^{+14}$ neutrons $cm^{-3}s^{-1}$ and $Q_{g,i,j} = 0$ for $(g,i,j) \neq (2,2,2)$ and $(3,2,2)$.

In order to verify wether or not this neutron source distribution drives the subcritical system to the prescribed power density ($P_{_{total}}$), we solved the forward four-group two-dimensional S$_{\text{8}}$ neutron transport equation corresponding to model problem I (\cref{2.2}). We used for this purpose the classical Diamond Difference (DD) method \cite{LeMi93}, considering a discretization grid composed of: $n_{x_{i}} = n_{y_{i}} = 88$ for $i = 1$ or $3$ and $n_{x_{2}} = n_{y_{2}} = 136$. \cref{t4,t5} display respectively the power generated in each fuel region and the node-average scalar fluxes as generated by the DD numerical solution of the forward four-group two-dimensional S$_{\text{8}}$ neutron transport equation. If we sum all values presented in \cref{t4} we obtain $P = 10.00007\,MW\,cm^{-1}$, which indicates that the offered methodology in fact determined the neutron source distribution with high accuracy. 
{\renewcommand{\arraystretch}{1.3}
	\begin{table}[t]
		\centering
		\caption{Power generated in each fuel region by the DD numerical solution of the forward four-group two-dimensional S$_{\text{8}}$ neutron transport problem for the first numerical experiment.}
		\small
		\begin{tabular}{c|ccc}
			\toprule[0.5mm]
			Region & \multicolumn{3}{c}{Power generated ($MW\,cm^{-1}$)} \\
			\cline{2-4}    ${R}_{_{i,j}}$ & $i = 1$ & $i = 2$ & $i = 3$  \\
			\hline
			$j = 3$ & 0.03546 & 1.40356 & 0.03546 \\
			\hline
			$j = 2$ & 1.40356 & 4.24399 & 1.40356 \\
			\hline
			$j = 1$ & 0.03546 & 1.40356 & 0.03546 \\
			\bottomrule[0.5mm]
		\end{tabular}%
		\label{t4}%
	\end{table}%
}
{\renewcommand{\arraystretch}{1.3}
	\begin{table}[t]
		\centering
		\caption{Node-average scalar fluxes generated by the DD numerical solution of the forward four-group two-dimensional S$_{\text{8}}$ neutron transport problem for the first numerical experiment.}
		\small
		\begin{threeparttable}
			\begin{tabular}{l|rrrrrr}
				\toprule[0.5mm]
				\multicolumn{1}{c|}{Node} & \multicolumn{6}{c}{Node-average scalar fluxes (neutrons $cm^{-2}s^{-1}$)} \\
				\cline{2-7}    \multicolumn{1}{c|}{$\Gamma_{_{k,\ell}}$} & \multicolumn{1}{c}{$\ell = 1$} & \multicolumn{1}{c}{$\ell = 62$} & \multicolumn{1}{c}{$\ell = 124$} & \multicolumn{1}{c}{$\ell = 186$} & \multicolumn{1}{c}{$\ell = 248$} & \multicolumn{1}{c}{$\ell = 312$} \\
				\hline
				 	$k=1$ & 2.63974E+16\tnote{a} & 3.35238E+16 & 8.02787E+16 & 8.18020E+16 & 3.43976E+16 & 2.63974E+16 \\
				\hline
				$k=62$ & 3.35238E+16 & 4.31612E+16 & 1.06163E+17 & 1.08040E+17 & 4.43358E+16 & 3.35238E+16 \\
				\hline
				$k=124$ & 8.02787E+16 & 1.06163E+17 & 1.79125E+17 & 1.81096E+17 & 1.09004E+17 & 8.02787E+16 \\
				\hline
				$k=186$ & 8.18020E+16 & 1.08040E+17 & 1.81096E+17 & 1.83060E+17 & 1.10905E+17 & 8.18020E+16 \\
				\hline
				$k=248$ & 3.43976E+16 & 4.43358E+16 & 1.09004E+17 & 1.10905E+17 & 4.55447E+16 & 3.43976E+16 \\
				\hline
				$k=312$ & 2.63974E+16 & 3.35238E+16 & 8.02787E+16 & 8.18020E+16 & 3.43976E+16 & 2.63974E+16 \\
				\bottomrule[0.5mm]
			\end{tabular}%
			\begin{tablenotes}[flushleft]
				\tiny
				\item[a] Read as 2.63974$\times10^{+16}$.
				\item[-] Node-average scalar fluxes summed in all energy groups.
			\end{tablenotes}
		\end{threeparttable}
		\label{t5}%
	\end{table}%
}

\subsubsection{Numerical experiment II}
For the second numerical experiment we suppose that the power density is distributed such that $30 \%$ $P_{_{total}}$ is generated by region $R_{_{2,2}}$ with the remaining $70\%$ $P_{_{total}}$ being produced by the other fuel regions. The linear system corresponding to this numerical experiment is displayed in \cref{t6}, where the row with respect to $P_{_{70}}$ is composed of the sum of all rows of the importance matrix built for model problem I, except the row associated with the power generated by region $R_{_{2,2}}$.
{\renewcommand{\arraystretch}{1.3}
	\begin{table}[t]{}
		\centering
		\caption{The linear system built for the second numerical experiment.}
		\small
		\begin{threeparttable}
\begin{tabular}{!{\vrule width 1pt}c!{\vrule width 1pt}c!{\vrule width 1pt}c:c:c:c:c:c!{\vrule width 1pt}c!{\vrule width 1pt}c!{\vrule width 1pt}}
 & \multirow{18}{*}{=} & 1,1,1  & 2,1,1  & 3,1,1  & 4,1,1  & 1,2,1  & 2,2,1 & \multirow{18}{*}{} & \multirow{18}{*}{$Q_{g,i,j}$}\tabularnewline
 &  & 6.89296E-16\tnote{c} \: & 4.55075E-16  & 1.29309E-16  & 8.74921E-17  & 1.57826E-15  & 1.5293E-15  &  & \tabularnewline
 &  & 1.46791E-15\tnote{d,e}$\,$ \: & 1.0348E-15  & 3.49488E-16  & 3.54772E-16  & 3,28793E-15  & 3.33858E-15  &  & \tabularnewline
\cline{3-8} \cline{4-8} \cline{5-8} \cline{6-8} \cline{7-8} \cline{8-8} 
 &  & 3,2,1  & 4,2,1  & 1,3,1  & 2,3,1  & 3,3,1  & 3,1,4  &  & \tabularnewline
 &  & 1.63238E-15  & 2.05489E-15  & 6.89296E-16  & 4.55075E-16  & 1.29309E-16  & 8.74921E-17  &  & \tabularnewline
 &  & 4,11753E-15  & 5.97368E-15  & 1.46791E-15  & 1.0348E-15  & 3.49488E-16  & 3.54772E-16  &  & \tabularnewline
\cline{3-8} \cline{4-8} \cline{5-8} \cline{6-8} \cline{7-8} \cline{8-8} 
 &  & 1,1,2  & 2,1,2  & 3,1,2  & 4,1,2  & 1,2,2  & 2,2,2  &  & \tabularnewline
 &  & 1.57825E-15  & 1.5293E-15  & 1.63238E-15  & 2.05488E-15  & 3.40328E-15  & 4.0331E-15  &  & \tabularnewline
$P_{_{2,2}}\tnote{a}\:\,$ &  & 3.28792E-15  & 3.33858E-15  & 4.11752E-15  & 5.97367E-15  & 5.5354E-15  & 5.86674E-15  &  & \tabularnewline
\cline{3-8} \cline{4-8} \cline{5-8} \cline{6-8} \cline{7-8} \cline{8-8} 
$P_{_{70}}\tnote{b} \:\,$  &  & 3,2,2  & 4,2,2  & 1,3,2  & 2,3,2  & 3,3,2  & 3,2,4  &  & \tabularnewline
 &  & 5.54179E-15  & 7.94425E-15  & 1.57825E-15  & 1.5293E-15  & 1.63238E-15  & 2.05488E-15  &  & \tabularnewline
 &  & 6.72270E-15  & 8.24923E-15  & 3.28792E-15  & 3.33858E-15  & 4.11752E-15  & 5.97367E-15  &  & \tabularnewline
\cline{3-8} \cline{4-8} \cline{5-8} \cline{6-8} \cline{7-8} \cline{8-8} 
 &  & 1,1,3 & 2,1,3  & 3,1,3  & 4,1,3  & 1,2,3  & 2,2,3  &  & \tabularnewline
 &  & 6.89296E-16  & 4.55075E-16  & 1.29309E-16  & 8.74921E-17  & 1.57826E-15  & 1.5293E-15  &  & \tabularnewline
 &  & 1.46791E-15  & 1.0348E-15  & 3.49488E-16  & 3.54772E-16  & 3.28793E-15  & 3.33859E-15  &  & \tabularnewline
\cline{3-8} \cline{4-8} \cline{5-8} \cline{6-8} \cline{7-8} \cline{8-8} 
 &  & 3,2,3  & 4,2,3  & 1,3,3  & 2,3,3  & 3,3,3  & 4,3,3  &  & \tabularnewline
 &  & 1.63238E-15  & 2.05489E-15  & 6.89296E-16  & 4.55075E-16  & 1.29309E-16  & 8.74921E-17  &  & \tabularnewline
 &  & 4.11753E-15  & 5.97368E-15  & 1.46791E-15  & 1.0348E-15  & 3.49488E-16  & 3.54772E-16  &  & \tabularnewline
\end{tabular}
			\begin{tablenotes}[flushleft]
				\tiny
				\item[a] First row of the importance matrix. This row is correlated with the power generated by region ${R}_{_{2,2}}$.
				\item[b]Second row of the importance matrix. This row is correlated with the power generated by the whole system except region ${R}_{_{2,2}}$.
			\item[c] Element $\left(a_{_{1,1}}\right)$ of the first row of the importance matrix that multiplies $Q_{1,1,1}$. The element index $1,z$ is defined such that $z = g + 4\left[ (i-1) + 3(j-1)\right]$.
				\item[d]Element $\left(a_{_{2,1}}\right)$ of the second row of the importance matrix that multiplies $Q_{1,1,1}$. The element index $2,z$ is defined such that $z = g + 4\left[ (i-1) + 3(j-1)\right] $.
				\item[e] Read as 1.46791$\times10^{-15}$.
				\item[-] Linear system of order $2\times 36$.
			\end{tablenotes}
		\end{threeparttable}
		\label{t6}%
	\end{table}%
}

To proceed, let us consider 34 auxiliary equations, in order to obtain a unique solution for this problem, 
\begin{equation}\label{e4.2}
Q_{g,i,j} = \left \{
\begin{array}{cl}
\mathcal{Q}_{1},& \text{if}\:\: (g,i,j) = (1,2,2)\\
\mathcal{Q}_{2},&  \text{for}\:\: i=1:3,\: j=1:3\:\: \text{and}\:g=1, \:\text{with}\: (g,i,j) \neq (1,2,2)\\
0,& \text{otherwise}
\end{array}
\right..
\end{equation}

Therefore, considering the auxiliary equations presented in \cref{e4.2}, we obtain the results: $\mathcal{Q}_{1}$ =  -4.41811$\times 10^{+14}$ neutrons $cm^{-3}s^{-1}$ and $\mathcal{Q}_{2}$ = 4.96527$\times 10^{+14}$ neutrons $cm^{-3}s^{-1}$. Clearly the neutron source distribution is physically incoherent, since negative sources do not make sense. This is an indication that our initial assumption for the power density and the considered auxiliary equations are not physically compatible. For the entire system, except region $R_{_{2,2}}$, to generate $70\%$ $P_{_{total}}$, it is necessary to introduce a source $\mathcal{Q}_{2}$ of a high intensity in the regions defined in \cref{e4.2}. However, neutrons within these regions can migrate to the adjacent regions and induce fission. Thus, the power generated in region $R_{_{2,2}}$ due to fission may be more intense than $30\%$ $P_{_{total}}$. Thus, a $``$negative source'' $\mathcal{Q}_{1}$ is required in order to remove neutrons from region $R_{_{2,2}}$, having this region produce $30\%$ $P_{total}$. To illustrate our assertive, we consider now a smoother power distribution, where $35\%$ $P_{_{total}}$ is generated by region $R_{_{2,2}}$ and $65\%$ $P_{_{total}}$ is produced by the remaining fuel regions. With this new power density distribution, we obtain the results: $\mathcal{Q}_{1}$ =  5.24640$\times 10^{+14}$ neutrons $cm^{-3}s^{-1}$ and $\mathcal{Q}_{2}$ = 1.89026$\times 10^{+14}$ neutrons $cm^{-3}s^{-1}$. Now no $``$negative sources'' has been generated; as a result, the power density and the auxiliary equations are physically compatible. We follow the same steps as in the first numerical experiment to validate the neutron source distribution. \cref{t7,t8} display respectively the power generated in each fuel region and the node-average scalar fluxes as generated by the corresponding forward four-group two-dimensional S$_{\text{8}}$ neutron transport problem.
{\renewcommand{\arraystretch}{1.3}
	\begin{table}[htbp]{}
		\centering
		\caption{Power generated in each fuel region by the DD numerical solution of the forward four-group two-dimensional S$_{\text{8}}$ neutron transport problem for the second numerical experiment.}
		\small
		\begin{tabular}{c|ccc}
			\toprule[0.5mm]
			Region & \multicolumn{3}{c}{Power generated ($MW\,cm^{-1}$)} \\
			\cline{2-4}    ${R}_{_{i,j}}$ & $i = 1$ & $i = 2$ & $i = 3$  \\
			\hline
			$j = 3$ & 0.04538 & 1.57956 & 0.04538 \\
			\hline
			$j = 2$ & 1.57956 & 3.50007 & 1.57956 \\
			\hline
			$j = 1$ & 0.04538 & 1.57956 & 0.04538 \\
			\bottomrule[0.5mm]
		\end{tabular}%
		\normalsize
		\label{t7}%
	\end{table}%
}{\renewcommand{\arraystretch}{1.3}
	\begin{table}[H]{}
		\centering
		\caption{Node-average scalar fluxes generated by the DD numerical solution of the four-group multigroup two-dimensional S$_{\text{8}}$ neutron transport problem for the second numerical experiment.}
		\small
		\begin{threeparttable}
			\begin{tabular}{l|rrrrrr}
				\toprule[0.5mm]
				\multicolumn{1}{c|}{Node} & \multicolumn{6}{c}{Node-average scalar fluxes (neutrons $cm^{-2}s^{-1}$)} \\
				\cline{2-7}    \multicolumn{1}{c|}{$\Gamma_{_{k,\ell}}$} & \multicolumn{1}{c}{$\ell = 1$} & \multicolumn{1}{c}{$\ell = 62$} & \multicolumn{1}{c}{$\ell = 124$} & \multicolumn{1}{c}{$\ell = 186$} & \multicolumn{1}{c}{$\ell = 248$} & \multicolumn{1}{c}{$\ell = 312$} \\
				\hline
				$k=1$ & 4.93487E+16\tnote{a} & 5.76182E+16 & 1.16376E+17 & 1.18197E+17 & 5.86839E+16 & 4.93487E+16 \\
				\hline
				$k=62$ & 5.76182E+16 & 6.64793E+16 & 1.30016E+17 & 1.31812E+17 & 6.76109E+16 & 5.76182E+16 \\
				\hline
				$k=124$ & 1.16376E+17 & 1.30016E+17 & 1.64635E+17 & 1.65691E+17 & 1.31444E+17 & 1.16376E+17 \\
				\hline
				$k=186$ & 1.18197E+17 & 1.31812E+17 & 1.65691E+17 & 1.66719E+17 & 1.33223E+17 & 1.18197E+17 \\
				\hline
				$k=248$ & 5.86839E+16 & 6.76109E+16 & 1.31444E+17 & 1.33223E+17 & 6.87476E+16 & 5.86839E+16 \\
				\hline
				$k=312$ & 4.93487E+16 & 5.76182E+16 & 1.16376E+17 & 1.18197E+17 & 5.86839E+16 & 4.93487E+16 \\
				\bottomrule[0.5mm]
			\end{tabular}%
			\begin{tablenotes}[flushleft]
				\tiny
				\item[a] Read as 4.93487$\times10^{+16}$.
				\item[-] Node-average scalar fluxes summed in all energy groups.
			\end{tablenotes}
		\end{threeparttable}
		\label{t8}%
		\normalsize
	\end{table}%
}

As we see in \cref{t7}, $P_{_{2,2}} = 3.50007\,MW\,cm^{-1}$, i.e., $P_{_{2,2}} \cong 0.35 \times P_{_{total}}$. Furthermore, if we sum all the values of power presented in \cref{t7} except $P_{_{2,2}}$, we obtain $P_{_{65}}= 6.49976\,MW\,cm^{-1}$, i.e., $P_{_{65}} \cong 0.65 \times P_{_{total}}$. 

\subsection{Model problem II}
For the second model problem we consider a square domain composed by 64 regions, 21 of which are composed of fuel materials. Figure \ref{f6} displays the structure of model problem II and \cref{t9} lists the material parameters used in this model problem. The discretization spatial grid in model problem II is defined such that $n_{x_{1}} = n_{y_{1}} = 14$ and $n_{x_{i}} = n_{y_{i}} = 28$ for $i=2:8$. Moreover, we used to model the adjoint problems the LQ$_{\text{4}}$ and the multigroup formulations, with $G = 2$ energy groups. With this configuration, the multiplication factor of the system is $k_{eff} = 0.98498$. Hence, the system is subcritical and tends to shutdown, unless fixed neutron sources are added to the system. We assume that the total power generated by the fuel regions per unit length of core height is $P_{_{total}} = 16\,MW\,cm^{-1}$.
\begin{figure}[]{}
	\centering
	\includegraphics[scale=0.25]{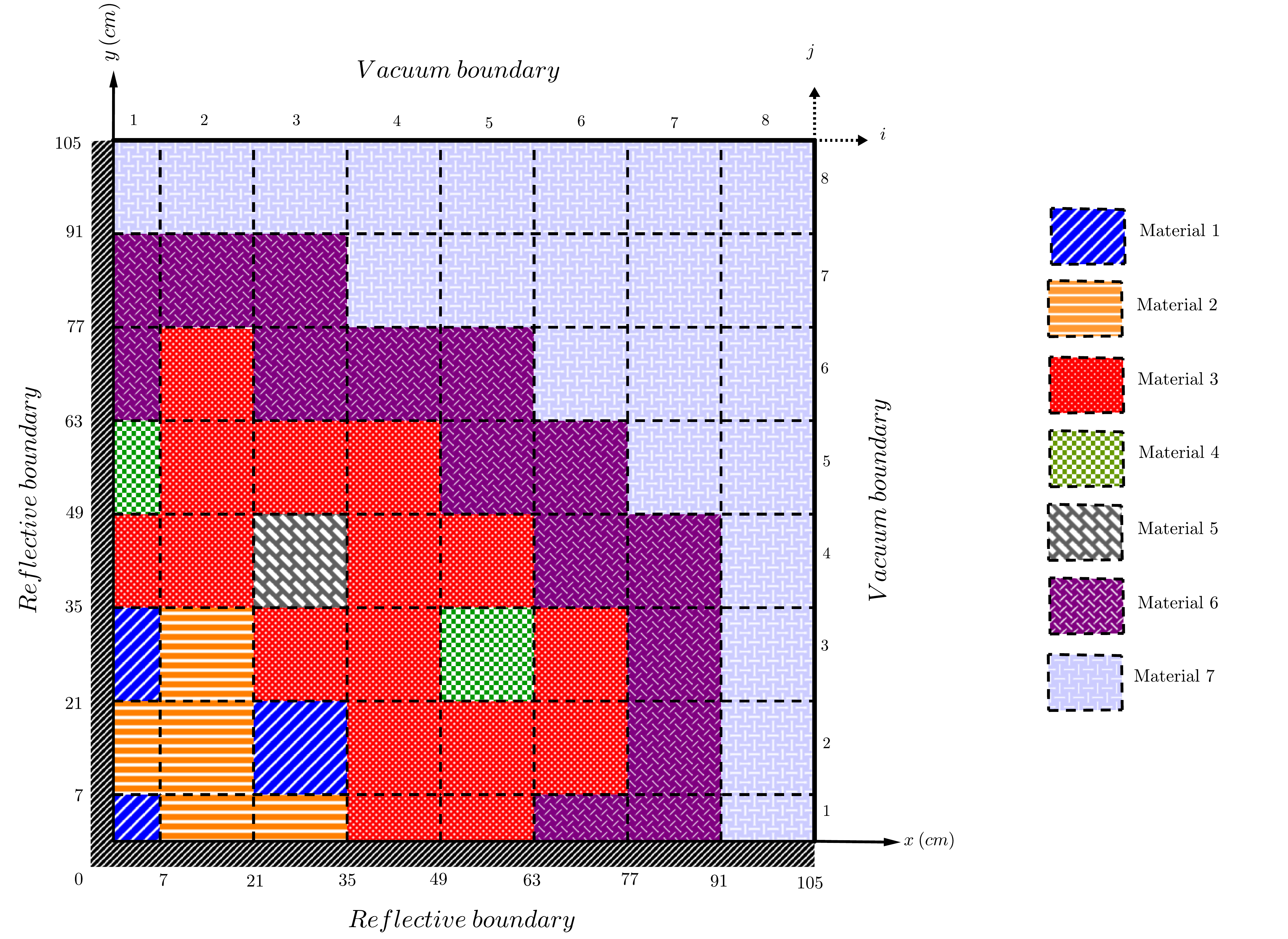}
		\caption{Model problem II.}
	\label{f6}
\end{figure} 
{\renewcommand{\arraystretch}{1.3}
	\begin{table}[htbp]{}
		\centering
		\caption{Material parameters of model problem II.}\label{t9}
		\small
		\begin{threeparttable}
			\begin{tabular}{cc|c|c|cc|c}
				\toprule[0,5mm]
				\multicolumn{2}{c|}{\multirow{2}[0]{*}{Material parameters}} & \multirow{2}[0]{*}{$\sigma_{t_{g}} (cm^{-1})$} & \multirow{2}[0]{*}{$\sigma_{f_{g}} (cm^{-1})$} & \multicolumn{2}{c|}{$\sigma_{s_{g\rightarrow g^{\prime}}} (cm^{-1})$} & \multirow{2}[0]{*}{$\chi_{g}$} \\
				\cline{5-6}    \multicolumn{2}{c|}{} &       &       & $g^{\prime} = 1$ & $g^{\prime} = 2$ &  \\
				\hline
				\multirow{2}[1]{*}{Material 1} & $g = 1$ & 2.31800E-01\tnote{a} & 0.00000E+00 & 1.82450E-01 & 1.56300E-02 &  \\
				& $g = 2$ & 8.38000E-01 & 0,00000E+00 & 0.00000E+00 & 7.33100E-01 &  \\
				\hline
				\multirow{2}[1]{*}{Material 2} & $g = 1$ & 2.31800E-01 & 2.36763E-03 & 1.84200E-01 & 1.63800E-02 & 1 \\
				& $g = 2$ & 8.71460E-01 & 5.16271E-02 & 0.00000E+00 & 7.42840E-01 & 0 \\
				\hline
				\multirow{2}[1]{*}{Material 3} & $g = 1$ & 2.36570E-01 & 3.06032E-03 & 2.09560E-01 & 1.45300E-02 & 1 \\
				& $g = 2$ & 8.22800E-01 & 6.95150E-02 & 0.00000E+00 & 7.06500E-01 & 0 \\
				\hline
				\multirow{2}[1]{*}{Material 4} & $g = 1$ & 2.71520E-02 & 0.00000E+00 & 2.64812E-02 & 1.63800E-04 &  \\
				& $g = 2$ & 5.22800E-02 & 0.00000E+00 & 0.00000E+00 & 5.10840E-02 &  \\
				\hline
				\multirow{2}[1]{*}{Material 5} & $g = 1$ & 2.22320E-01 & 0.00000E+00 & 2.15060E-01 & 1.15300E-03 &  \\
				& $g = 2$ & 8.17221E-01 & 0.00000E+00 & 0.00000E+00 & 7.26500E-01 &  \\
				\hline
				\multirow{2}[1]{*}{Material 6} & $g = 1$ & 3.37600E-01 & 0.00000E+00 & 3.03126E-01 & 1.01200E-03 &  \\
				& $g = 2$ & 9.99500E-01 & 0.00000E+00 & 0.00000E+00 & 3.13126E-01 &  \\
				\hline
				\multirow{2}[1]{*}{Material 7} & $g = 1$ & 1.78120E-01 & 0.00000E+00 & 8.00710E-02 & 3.43400E-03 &  \\
				& $g = 2$ & 1.17616E+00 & 0.00000E+00 & 0.00000E+00 & 1.14458E-01 &  \\
				\bottomrule[0,5mm]
			\end{tabular}%
			\normalsize
			\begin{tablenotes}[flushleft]
				\tiny
				\item[a] Read as 2.31800$\times10^{-01}$.
			\end{tablenotes}
		\end{threeparttable}
	\end{table}
}

In order to determine the neutron source distributions we first build the importance matrix correlated with this model problem. Therefore, we solve \cref{2.5} 21 times using sequentially the adjoint source term according to \cref{e2.7}. Figure \ref{f7} illustrates the node-average adjoint scalar fluxes for $i^{\star},j^{\star} = 4,3$ (\cref{e2.7}) as a function of the node midpoints.
\begin{figure}[H]
	\begin{subfigure}{.5\textwidth}
		\centering
		\includegraphics[scale=0.5]{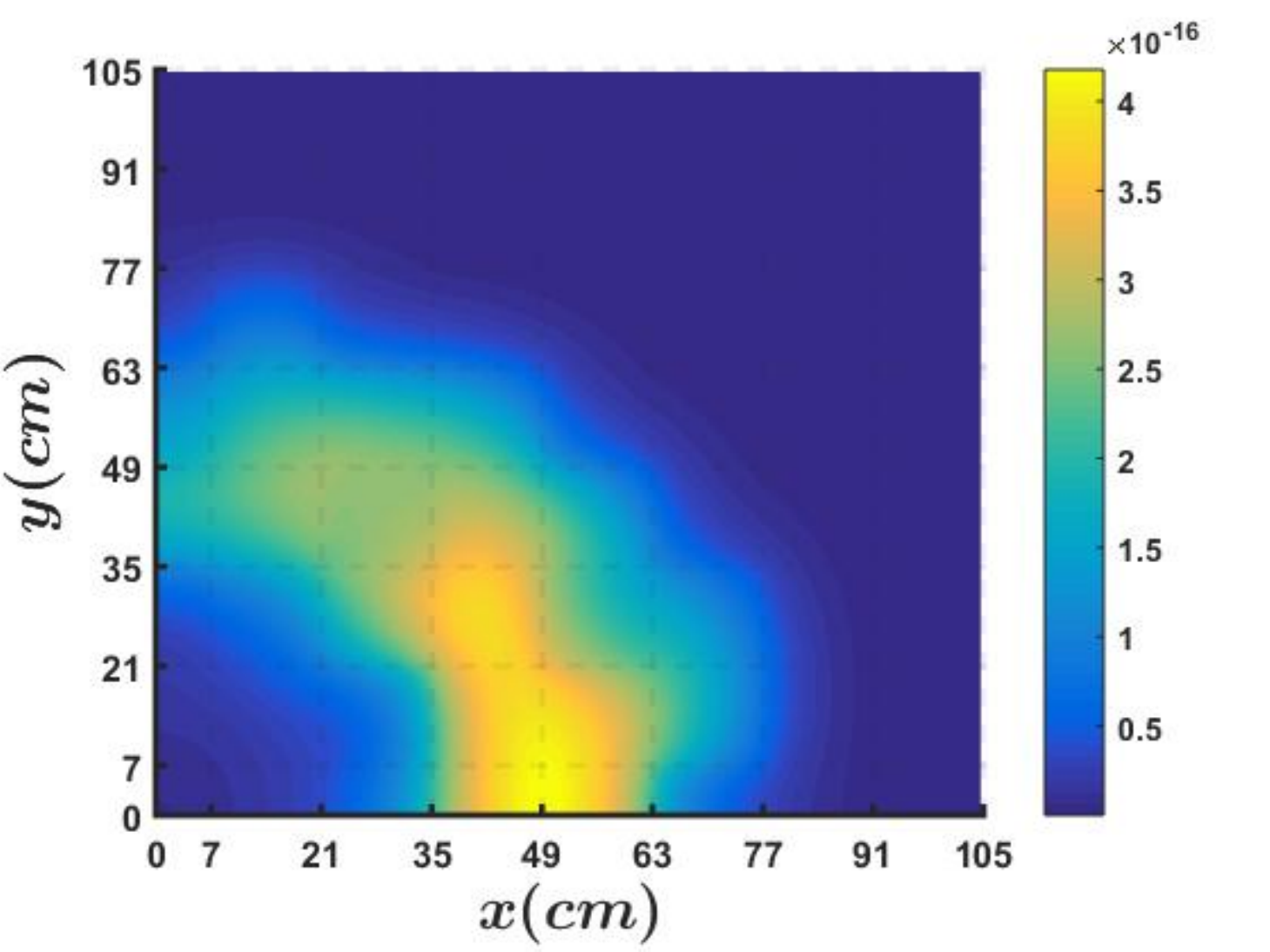}
		\caption{$g = 1$}
	\end{subfigure}
	\begin{subfigure}{.5\textwidth}
		\centering
		\includegraphics[scale=0.5]{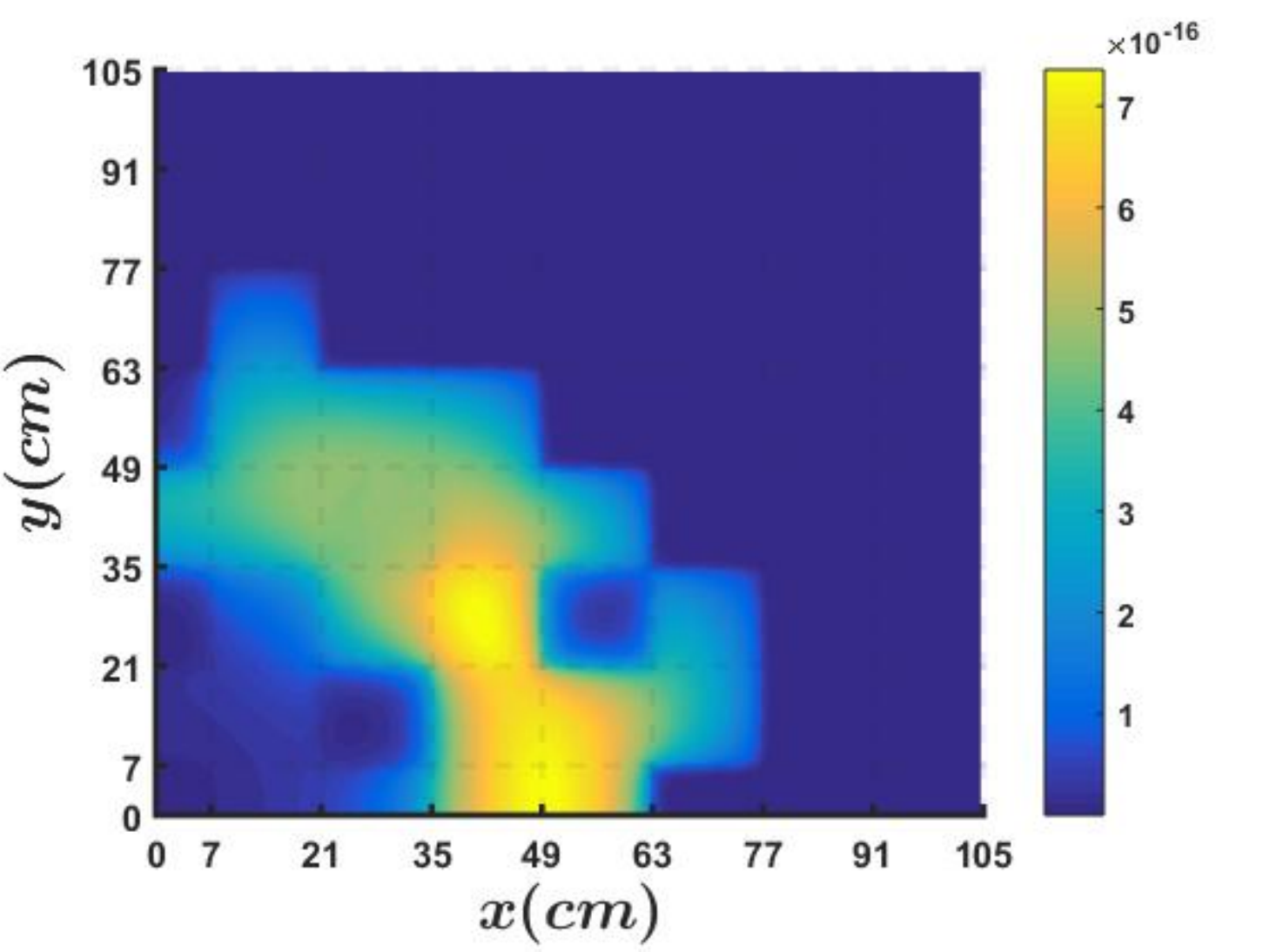}
		\caption{$g = 2$}
	\end{subfigure}
	\caption{Node-average adjoint scalar fluxes considering the adjoint source term with $i^{\star},j^{\star} = 4,3$ for model problem II.}
	\label{f7}
\end{figure}

\subsection{Numerical experiment III}
In the third numerical experiment we assume that the power density is distributed such that the regions composed of material 2 generate $20\%$ $P_{_{total}}$ and the regions composed of material 3 produce the remaining $80\%$ $P_{_{total}}$. Since the importance matrix for model problem II has been already built, we follow two distinct procedures: (i) we sum the rows of the importance matrix correlated to the power generated by regions composed of material 2; and (ii) sum the rows correlated to the power generated by regions composed of material 3. By doing this, we obtain the linear system of 2 equations and 128 unknowns. 

Once the linear system is defined, we need to provide 126 auxiliary equations in order to obtain a unique solution for the problem. Thus, we consider
\begin{equation}\label{e4.3}
Q_{g,i,j} = \left \{
\begin{array}{cl}
\mathcal{Q}_{1},& \text{if}\:\: (g,i,j) = (1,1,1)\\ 
0.005\times \mathcal{Q}_{1},&  \text{if}\:\: (g,i,j) = (2,1,1)\\
\mathcal{Q}_{2},& \text{if}\:\: (g,i,j) = (1,3,1) \text{ or } (1,3,2) \\[-0.1cm]
0.005\times \mathcal{Q}_{2},&  \text{if}\:\: (g,i,j) = (2,3,1) \text{ or } (2,3,2)\\
0,& \text{otherwise}
\end{array}
\right..
\end{equation}

{\renewcommand{\arraystretch}{1.1}
	\begin{table}[htbp]{}
		\centering
		\caption{The linear system built for the third numerical experiment.}
		\small
		\begin{threeparttable}
		\begin{tabular}{!{\vrule width 1pt}c!{\vrule width 1pt}c!{\vrule width 1pt}c:c:c!{\vrule width 1pt}c!{\vrule width 1pt}c!{\vrule width 1pt}}
 & \multirow{6}{*}{=} & 1,1,1  & 2,1,1  & 1,3,1 & \multirow{6}{*}{} & \multirow{6}{*}{$Q_{g,i,j}$}\tabularnewline
 &  & 1.31488E-15\tnote{c}\,\:  & 1.19912E-15  & 4.16063E-15  &  & \tabularnewline
$P_{_{20}}\tnote{a}\,\:$ &  & 4.02961E-15\tnote{d,e}$\:\:\:\:$  & 1.81302E-15  & 5.33428E-14  &  & \tabularnewline
\cline{3-5} \cline{4-5} \cline{5-5} 
$P_{_{80}}\tnote{b}\,\:$ &  & 2,3,1  & 1,3,2  & 2,3,2  &  & \tabularnewline
 &  & 3.08336E-15  & 1,13206E-14  & 7,85309E-15  &  & \tabularnewline
 &  & 3.03778E-14  & 1.91951E-13  & 1.19016E-13  &  & \tabularnewline
\end{tabular}
			\begin{tablenotes}[flushleft]
				\tiny
				\item[a]First row of the importance matrix. This row is correlated to the power generated by regions composed of material 2.
				\item[b]Second row of the importance matrix. This row is correlated to the power generated by regions composed of material 3.
				\item[c]  Element $\left(a_{_{1,1}}\right)$ of the first row of the importance matrix that multiplies $Q_{1,1,1}$. The element index $1,z$ is defined such that $z = g + 2\left[ (i-1) + 8(j-1)\right]$.
				\item[d] Element $\left(a_{_{2,1}}\right)$ of the second row of the importance matrix that multiplies $Q_{1,1,1}$. The element index $2,z$ is defined such that $z = g + 2\left[ (i-1) + 8(j-1)\right]$.
				\item[e] Read as 4.02961$\times 10^{-15}$.
				\item[-] Linear system of order $2\times 128$.
			\end{tablenotes}
		\end{threeparttable}
		\label{t10}%
\end{table}}

According to \cref{e4.3}, the linear system as presented in \cref{t10} does represent this numerical experiment. Therefore, we obtain: $\mathcal{Q}_{1} = $ 2.24396$\times 10^{+15}$ neutrons $cm^{-3}s$ and $\mathcal{Q}_{2} = $ 1.51898$\times 10^{+13}$ neutrons $cm^{-3}s^{-1}$. 

Now, in order to verify wether of not this neutron source distribution drives the subcritical system to the prescribed power density, we follow the same steps as in model problem I. Thus, we solve the forward two-group S$_{\text{4}}$ neutron transport problem, making use of the DD method, considering a discretization spatial grid composed of: $n_{x_{1}} = n_{y_{1}} = 35$ and $n_{x_{i}} = n_{y_{i}} = 70$ for $i = 2:8$. \cref{t11,t12} display respectively the power generated in each fuel region and the node-average scalar fluxes as generated by the forward two-group two-dimensional S$_{\text{4}}$ neutron transport problem.
{\renewcommand{\arraystretch}{1.3}
	\begin{table}[htbp]{}
		\centering
		\caption{Power generated in each fuel region by the DD numerical solution of the forward two-group two-dimensional S$_{\text{4}}$ neutron transport problem for the third numerical experiment.}
		\small
		\begin{threeparttable}
			\begin{tabular}{c|cccccccc}
				\toprule[0,5mm]
				Region & \multicolumn{8}{c}{Power generated ($MW\,cm^{-1}$)} \\
				\cline{2-9}    ${R}_{_{i,j}}$ & $i= 1$ & $i = 2$ & $i= 3$ & $i = 4$ & $i = 5$ & $i = 6$ & $i = 7$ & $i = 8$ \\
				\hline
				$j= 8$ &  &  &  &  &  &  &  & \\
				\hline
				$j = 7$ &  &  &  &  &  &  &  & \\
				\hline
				$j = 6$ &  & 0.28432 &  &  &  &  &  &  \\
				\hline
				$j = 5$ &  & 0.80543 & 0.82001 & 0.59259 &  &  &  &   \\
				\hline
				$j = 4$ & 0.40215 & 0.97907 & & 1.09033 & 0.56148 &  &  &  \\
				\hline
				$j = 3$ &  & 0.45559 & 1.00168 & 1.24914 &  & 0.41679 &  &  \\
				\hline
				$j = 2$ & 0.83056 & 0.81729 &  & 1.34847 & 1.27299 & 0.58111 &  & \\
				\hline
				$j = 1$ &  & 0.83956 & 0.25697 & 0.71200 & 0.68247 &  &  &  \\
				\bottomrule[0,5mm]
			\end{tabular}%
			\begin{tablenotes}[flushleft]
				\tiny
				\item[-] In the blank spaces the power generated is equal to zero.
			\end{tablenotes}
		\end{threeparttable}
		\normalsize
		\label{t11}%
	\end{table}%
}{\renewcommand{\arraystretch}{1.3}
	\begin{table}[htbp]{}
		\centering
		\caption{Node-average scalar fluxes generated by the DD solution of the forward multigroup two-dimensional S$_{\text{4}}$ neutron transport problem for the third numerical experiment.}
		\small
		\begin{threeparttable}
			\begin{tabular}{l|cccccc}
				\toprule[0,5mm]
				\multicolumn{1}{c|}{Node} & \multicolumn{6}{c}{Node-average scalar fluxes (neutrons $cm^{2}s^{-1}$)} \\
				\cline{2-7}    \multicolumn{1}{c|}{$\Gamma_{_{k,\ell}}$} & $\ell = 1$ & $\ell = 106$ & $\ell = 211$ & $\ell = 316$ & $\ell = 421$ & $\ell = 525$ \\
				\hline
				$k=1$ & 1.33808E+17\tnote{a} & 1.05594E+16 & 1.33762E+16 & 4.95777E+15 & 2.95080E+14 & 3.10090E+12 \\
				\hline
				$k=106$ & 1.22621E+16 & 8.64269E+15 & 1.67456E+16 & 8.10275E+15 & 3.42656E+14 & 2.48345E+12 \\
				\hline
				$k=211$ & 2.31758E+16 & 2.12500E+16 & 1.76970E+16 & 4.48211E+15 & 4.63146E+13 & 7.60867E+11 \\
				\hline
				$k=316$ & 1.19779E+16 & 1.29534E+16 & 4.65335E+15 & 2.64865E+14 & 6.40438E+12 & 2.49957E+11 \\
				\hline
				$k=421$ & 6.10849E+14 & 9.83244E+14 & 2.22748E+14 & 8.89622E+12 & 6.73789E+11 & 1.37654E+10 \\
				\hline
				$k=525$ & 6.67356E+12 & 6.95179E+12 & 2.54976E+12 & 3.53132E+11 & 1.55609E+10 & 9.25709E+10 \\
				\bottomrule[0,5mm]
			\end{tabular}%
			\begin{tablenotes}[flushleft]
				\tiny
				\item[a]Read as 1.33808$\times10^{+17}$.
				\item[-] Node-average scalar fluxes summed in all energy groups.
			\end{tablenotes}
		\end{threeparttable}
		\label{t12}%
	\end{table}%
}

According to \cref{t11}, if we sum the power generated by the regions composed of material 2 we obtain $P_{_{20}} = 3.19997 \,MW\,cm^{-1}$, i.e., $P_{_{20}} \cong 0.2\times P_{_{total}}$ and if we sum the power generated by regions composed of material 3 we obtain $P_{_{80}} = 12.80003 \,MW\,cm^{-1}$, i.e., $P_{_{80}} \cong 0.8\times P_{_{total}}$. Therefore, the offered methodology determined the neutron source distribution with high accuracy.

\section{Discussion}\label{sec5}
Motivated by the principle of operation of ADS devices, we presented in this work a methodology that allows one to determine the neutron source distribution which must be added into a subcritical system to drive it to a steady-stade prescribed power density level. The methodology is based on a relation between a linear functional with respect to the neutron angular flux, and the importance function associated with this functional.
The functional represents in the present two-dimensional S$_{\text{N}}$ model the power generated by a fuel region per unit length of the core height and the associated importance function measures the contribution that a neutron inserted into the system has to this generation of power.

 The main point of the offered technique consists in determining the importance matrix, which is composed of solutions of the energy multigroup two-dimensional adjoint S$_{\text{N}}$ equations. Each solution of these adjoint equations is associated with the power generated by one fuel region; therefore, a linear system is built using the importance matrix to correlate the interior neutron sources and the power generated by the subcritical system. At this point we remark that the importance matrix associated to a specific problem is built only one time, no matter how many numerical experiments are to be considered by the user, since any distribution of power is fundamentally a linear relation of the power generated by the fuel regions individually. Therefore, the importance matrix associated with any numerical experiment, as defined by the user, can be further built after the importance matrix correlated with the problem considered. For problems with high number of fuel regions the process of calculating the importance matrix can become time consuming, since it is required to solve the adjoint equations $I^{\star}J^{\star}$ times, being $I^{\star}J^{\star}$ the total number of fuel regions. However, this algorithm can run in parallel, since the adjoint equations for different adjoint source terms are not coupled.

In order to analyze the accuracy and characteristics of the offered methodology we presented three numerical experiments with respect to two model problems in the previous section. From those results we conclude that the offered methodology determined the neutron source distribution with very high accuracy. In fact, the more accurate the solutions of the adjoint equations, the more accurately the present methodology is expected to determine the neutron source distribution. Should the analytical solution of two-dimensional adjoint problem be available, the offered methodology would determine the exact values for the neutron source distribution, apart from rounding errors due to finite computational arithmetic. Therefore, as future work, a sensitivity analysis  is of interest to quantity the influence of the numerical methods used for solving \SN adjoint problems on the accuracy of the neutron source distribution.

An interesting point about the importance function being correlated with the power generation lies in its ability to help decision making.  For example, in the context of the ADS devices, as the importance function provides information about the expected contribution of a neutron inserted into the system to the generation of power, different configurations for the system will consequently generate different profiles for the importance functions. Thus, the configuration which produces the highest global value for the importance function should probably lead to less intense fixed neutron sources in order to drive the system to a prescribed power level.

We intend to continue this work by extending the mathematical model adopted here to three-dimensional calculations. To achieve this goal, we first suggest the investigation of the application of parallel techniques for the computer programs in the following steps of the algorithm: (i) the solution of the eigenvalue problem, cf. \cref{e3.5}, in each material zone; (ii) construction of the matrices defined in the RM$^{\dagger}$-CN discretized equations (\cref{e3.14a,e3.15a}); (iii) partial NBI$^{\dagger}$ sweeping matrices; and (iv) generation of the importance matrix. We expect this to improve the efficiency of the computer code and we shall report on the results when they are fully tested.

\section*{Acknowledgements}
This study was financed in part by the Coordenação de Aperfeiçoamento de Pessoal de Nível Superior - Brasil (CAPES) - Finance Code 001. The authors also would like to express their gratitude to the support of Conselho Nacional de Desenvolvimento Científico e Tecnológico - Brasil (CNPq) - and Fundação Carlos Chagas Filho de Amparo à Pesquisa do Estado do Rio de Janeiro - Brasil (FAPERJ).

\bibliography{Notas}

\end{document}